\documentclass[aps,pra,preprint,groupedaddress,showpacs,showkeys]{revtex4}

\usepackage {graphicx}
\usepackage {tabularx}
\usepackage{color}
\usepackage {ulem}
\usepackage{amsmath}

\begin{document}


\title{Basic variables to be reproduced in the first-principles theory for 
superconductors: Fluctuation of the particle number}



\author{Katsuhiko Higuchi}
\affiliation{Graduate School of Advanced Sciences of Matter, 
Hiroshima University, Higashi-Hiroshima 739-8527, Japan}

\author{Masahiko Higuchi}
\affiliation{Department of Physics, Faculty of Science, Shinshu University, 
Matsumoto 390-8621, Japan}


\date{\today}

\begin{abstract}
We show that the diagonal elements of the second-order reduced density 
matrix (RDM2) can be chosen as basic variables for describing the 
superconducting state, instead of the off-diagonal elements of the RDM2 that 
are usually adopted as basic variables in the density functional scheme. The 
diagonal elements of the RDM2 are called pair-density (PD), which is 
explicitly related to the fluctuation of the particle number of the system. 
In this paper, we argue that the fluctuation of the particle number can 
become an indication of the superconducting state, and that the density 
functional scheme in which the PD is chosen as a basic variable would be a 
promising first-principles theory for superconductors.
\end{abstract}


\maketitle


\section{\label{secI}Introduction}
The superconducting state of the fermion system is defined as the 
condensation of electrons into the same two-particle state, which is 
so-called the Bose-Einstein condensation (BEC) of the fermion system \cite{1}. 
According to this definition, the off-diagonal elements of the second-order 
reduced density matrix (RDM2) of the system is related to the order 
parameter of the superconducting state (OPSS) \cite{1,2,3,4}. The OPSS is 
sometimes called the off-diagonal long-range order \cite{2,3,4}. The OPSS 
explicitly includes the two-particle wave function which forms the BEC, and 
accordingly reflects its spatial and spin symmetries rigorously \cite{2,3,5}. 
Since these information are useful for understanding the properties of the 
superconductor, the first-principles theories have been developed for the 
purpose of calculating directly the OPSS \cite{6,7,8,9,10}. Especially in the 
field of the density functional theory (DFT), the research has been active 
since the first theory presented by Oliveira, Gross and Kohn \cite{9,10}. A lot 
of works including the actual applications to various superconductors 
\cite{11,12,13,14,15,16,17,18,19,20,21,22,23,24,25,26,27,28,29} 
have been done so far. In all these works, the OPSS, i.e., 
off-diagonal elements of the RDM2, are certainly adopted as the basic 
variables that describe the equilibrium properties of the superconducting state 
\cite{11,12,13,14,15,16,17,18,19,20,21,22,23,24,25,26,27,28,29}.

Thus the previous DFT-based theories for superconductors focuses on the off-diagonal 
elements of the RDM2 \cite{11,12,13,14,15,16,17,18,19,20,21,22,23,24,25,26,27,28,29}. 
On the other hand, as shown in this paper, there exists the mutual relation between 
the off-diagonal elements of the RDM2 and the diagonal ones in the superconducting state. 
In other words, the off-diagonal elements of the RDM2 and the diagonal ones are 
not independent of each other in the superconducting state. This means that 
it is not necessary to use the off-diagonal elements of the RDM2 so as to 
describe the superconducting state, but that it is possible to use the 
diagonal elements instead of them. The diagonal elements of the RDM2 are 
called the pair-density (PD), which is explicitly related to the fluctuation 
of the particle number (Sec. IV). In this paper we will show that the 
fluctuation of the particle number, i.e., diagonal elements of the RDM2, can 
become an indication of whether the superconducting state appears or not in 
the system.

As a promising theory to predict the fluctuation of the particle number, we come up with 
the PD functional theory \cite{30,31,32,33,34,35,36,37,38,39,40,41,42,43,44,45,46,47,48,49,
50,51,52,53,54,55,56,57,58,59,60}. 
This is because the fluctuation of the particle number is calculated rigorously with 
the use of the PD and electron density, which will also be shown in the subsequent 
sections. In addition to the above reason, there would be an advantageous 
reason of developing the PD functional theory from another point of view. 
That is, it can be expected to utilize the findings and knowledge obtained 
in developing zero-temperature PD functional theory for the normal state 
material \cite{30,31,32,33,34,35,36,37,38,39,40,41,42,43,44,45,46,47,48,49,
50,51,52,53,54,55,56,57,58,59,60}. 
The zero-temperature PD functional theory 
has recently attracted much attention due to the reason that the PD contains 
full of the information on the static electron correlation \cite{32}. Various 
approaches have attempted to overcome or avoid the problems of 
zero-temperature PD functional theory \cite{30,31,32,33,34,35,36,37,38,39,40,41,
42,43,44,45,46,47,48,49,50,51,52,53,54,55,56,57,58,59,60}. Especially, 
so-called the ``scaling method'' which is an effective method of searching 
PDs \cite{37,39}, and the coupling-constant expression for the kinetic energy 
functional \cite{38,40} seem to be useful for the case of the superconductor. The 
finite-temperature PD functional theory for superconductors, which can 
predict the fluctuation of the particle number, has not yet proposed so far. 
In this paper, we present the theoretical framework of such the PD 
functional theory.

Organization of this paper is as follows. The superconducting state is 
defined as the BEC of the fermion system. Using this definition, the OPSS is 
reviewed in Sec. II for the convenience of the subsequent discussions. In 
Sec. III, we show that the fluctuation of the particle number becomes $O(N)$ 
when the OPSS appears in the system. This means that there exists another 
possible way to choose basic variables for describing the superconducting 
state. In order to develop the first-principles theory for reproducing the 
fluctuation of the particle number, in Sec. IV, we present the PD functional 
theory that is applicable to the superconductor. In Sec. V, some concluding 
remarks are presented together with the discussions on the signification of 
this PD functional theory.
%
\section{Order parameter of the superconducting state}
\label{secII}
The superconducting state is defined with the aid of the concept of the BEC 
of the interacting fermion system \cite{1}. In this section, we shall review the 
OPSS \cite{2,3,4} with emphasize on the spatial broadenings and occupation 
numbers of two-particle states which form the BEC. These are essential for 
the definition of the OPSS. 
%
\subsection{Spectrum decomposition of the RDM2}
\label{secII-A}
In order to define the superconducting state, it is necessary to consider 
two-particle states which are included in the RDM2 of the many-electron 
system. The operator of the RDM2 is given by \cite{61}
\begin{equation}
\label{eq1}
\hat{{D}}^{(2)}\left( {{\rm {\bf r}}_{1} \zeta_{1} ,{\rm {\bf r}}_{2} \zeta 
_{2} ;{\rm {\bf {r}'}}_{1} {\zeta }'_{1} ,{\rm {\bf {r}'}}_{2} {\zeta }'_{2} 
} \right)={\frac{1}{2}}\psi^{\dag }({\rm {\bf {r}'}}_{1} {\zeta }'_{1} 
)\psi^{\dag }({\rm {\bf {r}'}}_{2} {\zeta }'_{2} )\psi ({\rm {\bf r}}_{2} 
\zeta_{2} )\psi ({\rm {\bf r}}_{1} \zeta_{1} ),
\end{equation}
where $\psi ({\rm {\bf r}}_{1} \zeta_{1} )$ and $\psi^{\dag }({\rm {\bf 
r}}_{1} \zeta_{1} )$ are field operators of electrons, and ${\rm {\bf r}}$ 
and $\zeta $ are spatial and spin coordinates, respectively. If the RDM2 of 
the system of state ${\left| {\Psi} \right\rangle }$ is denoted as $\left\langle 
{\hat{{D}}^{(2)}\left( {{\rm {\bf r}}_{1} \zeta_{1} ,{\rm {\bf r}}_{2} 
\zeta_{2} ;{\rm {\bf {r}'}}_{1} {\zeta }'_{1} ,{\rm {\bf {r}'}}_{2} {\zeta 
}'_{2} } \right)} \right\rangle_{\Psi } $, it is given by
\begin{eqnarray}
\label{eq2}
 \left\langle {\hat{{D}}^{(2)}\left( {{\rm {\bf r}}_{1} \zeta_{1} ,{\rm 
{\bf r}}_{2} \zeta_{2} ;{\rm {\bf {r}'}}_{1} {\zeta }'_{1} ,{\rm {\bf 
{r}'}}_{2} {\zeta }'_{2} } \right)} \right\rangle_{\Psi } 
&=& {\left\langle 
{\Psi } \right|}\hat{{D}}^{(2)}\left( {{\rm {\bf r}}_{1} \zeta_{1} ,{\rm 
{\bf r}}_{2} \zeta_{2} ;{\rm {\bf {r}'}}_{1} {\zeta }'_{1} ,{\rm {\bf 
{r}'}}_{2} {\zeta }'_{2} } \right){\left| {\Psi} \right\rangle }  \nonumber \\
&=& {\frac{1}{2}}{\left\langle {\Psi } \right|}\psi^{\dag }({\rm {\bf 
{r}'}}_{1} {\zeta }'_{1} )\psi^{\dag }({\rm {\bf {r}'}}_{2} {\zeta }'_{2} 
)\psi ({\rm {\bf r}}_{2} \zeta_{2} )\psi ({\rm {\bf r}}_{1} \zeta_{1} 
){\left| {\Psi} \right\rangle }, 
\end{eqnarray}
Using Eq. (\ref{eq2}), we shall define the operator $\hat{{D}}_{\Psi }^{(2)} $ as 
follows:
\begin{equation}
\label{eq3}
{\left\langle {{\rm {\bf r}}_{1} \zeta_{1} ,{\rm {\bf r}}_{2} \zeta_{2} } 
\right|}\hat{{D}}_{\Psi }^{(2)} {\left| {{\rm {\bf {r}'}}_{1} {\zeta }'_{1} ,{\rm {\bf 
{r}'}}_{2} {\zeta }'_{2} } \right\rangle}
= 2\left\langle {\hat{{D}}^{(2)}\left( {{\rm {\bf r}}_{1} \zeta_{1},{\rm 
{\bf r}}_{2} \zeta_{2} ;{\rm {\bf {r}'}}_{1} {\zeta }'_{1} ,{\rm {\bf 
{r}'}}_{2} {\zeta }'_{2} } \right)} \right\rangle_{\Psi },
\end{equation}
where, as $\left| {0} \right\rangle$ is the vacuum state, the ket vector 
${\left|  {{\rm {\bf r}}_{1} \zeta_{1} ,{\rm {\bf r}}_{2} \zeta_{2} }  \right\rangle }$ is given by 
\begin{equation}
\label{eq4}
{\left|  {{\rm {\bf r}}_{1} \zeta_{1} ,{\rm {\bf r}}_{2} \zeta_{2} }  \right\rangle }
=\psi^{\dag }({\rm {\bf {r}}}_{1} {\zeta }_{1} )\psi^{\dag }({\rm {\bf {r}}}_{2} {\zeta }_{2} )
\left| {0} \right\rangle.
\end{equation}
If the eigenfunction and eigenvalue for the operator $\hat{{D}}_{\Psi}^{(2)} $ 
are denoted as $\left| {\nu}_{\Psi} \right\rangle$ and 
$n_{\nu_{\Psi } }^{(2)} $, respectively, the eigenvalue equation for 
$\hat{{D}}_{\Psi }^{(2)} $ is expressed as
\begin{equation}
\label{eq5}
\hat{{D}}_{\Psi }^{(2)} \left| {\nu}_{\Psi} \right\rangle 
=n_{\nu_{\Psi } }^{(2)} \left| {\nu}_{\Psi} \right\rangle. 
\end{equation}
The eigenfunctions $\left| {\nu}_{\Psi} \right\rangle$ are two-particle 
states, which can be formally defined corresponding to the many-electron 
state ${\left| {\Psi} \right\rangle }$. They are sometimes called geminal in the 
field of the quantum chemistry \cite{61}. We suppose that they form the 
orthonormal and complete set. Using these eigenfunctions and eigenvalues, 
the spectrum decomposition of the RDM2 is written as
\begin{equation}
\label{eq6}
\left\langle {\hat{{D}}^{(2)}\left( {{\rm {\bf r}}_{1} \zeta_{1} ,{\rm {\bf 
r}}_{2} \zeta_{2} ;{\rm {\bf {r}'}}_{1} {\zeta }'_{1} ,{\rm {\bf {r}'}}_{2} 
{\zeta }'_{2} } \right)} \right\rangle_{\Psi } =\sum\limits_{\nu_{\Psi } } 
{n_{\nu_{\Psi } }^{(2)} } \nu_{\Psi } \left( {{\rm {\bf r}}_{1} \zeta_{1} 
,{\rm {\bf r}}_{2} \zeta_{2} } \right)\nu_{\Psi } \left( {{\rm {\bf 
{r}'}}_{1} {\zeta }'_{1} ,{\rm {\bf {r}'}}_{2} {\zeta }'_{2}} \right)^{\ast },
\end{equation}
where $\nu_{\Psi } \left( {{\rm {\bf r}}_{1} \zeta_{1} ,{\rm {\bf r}}_{2} 
\zeta_{2} } \right)$ is the coordinate representation of two-particle state 
${\left| {\nu_{\Psi}} \right\rangle }$, which is given by $\nu_{\Psi } 
\left( {{\rm {\bf r}}_{1} \zeta_{1} ,{\rm {\bf r}}_{2} \zeta_{2} } \right)
={\left\langle {{\rm {\bf r}}_{1} \zeta_{1} ,{\rm {\bf r}}_{2} \zeta _{2} } 
\left| \right.  {\nu_{\Psi } } \right\rangle } /{\sqrt 2 }$. 
The spatial broadening of $\nu_{\Psi } \left( {{\rm {\bf 
r}}_{1} \zeta_{1} ,{\rm {\bf r}}_{2} \zeta_{2} } \right)$ is intimately 
associated with whether the system contains the pairing state such as the 
Cooper pair of the superconducting state or not. The magnitude of $n_{\nu 
_{\Psi } }^{(2)} $ is also significant for judging whether the 
superconducting state appears or not, because $n_{\nu_{\Psi } }^{(2)} $ 
corresponds to the occupation number of two-particle state $\nu_{\Psi } 
\left( {{\rm {\bf r}}_{1} \zeta_{1} ,{\rm {\bf r}}_{2} \zeta_{2} } 
\right)$ \cite{2}. The maximum value of $n_{\nu_{\Psi } }^{(2)} $ is also shown 
to be not $O(N^{2})$ but $O(N)$\cite{1}. As will be discussed in the following 
subsection, both the spatial broadening of $\nu_{\Psi } \left( {{\rm {\bf 
r}}_{1} \zeta_{1} ,{\rm {\bf r}}_{2} \zeta_{2} } \right)$ and the 
magnitude of $n_{\nu_{\Psi } }^{(2)} $ are key quantities for the 
definition of the superconducting state. 
%
\subsection{Definition of the superconductivity}
\label{secII-B}
Using the coordinate of center of gravity ${\rm {\bf R}}$ and relative 
coordinate ${\boldsymbol \rho}$, where 
${\rm {\bf R}}={( {\rm {\bf r}}_{1} +{\rm {\bf r}}_{2} )}/2$, 
$\boldsymbol \rho={\rm {\bf r}}_{1} -{\rm {\bf r}}_{2}$, 
two-particle state (geminal) is rewritten as 
$\nu _{\Psi } \left( {{\rm {\bf R}{\boldsymbol \rho}};\zeta_{1} \zeta_{2} } \right)$ instead 
of $\nu_{\Psi } \left( {{\rm {\bf r}}_{1} \zeta_{1} ,{\rm {\bf r}}_{2} 
\zeta_{2} } \right)$. Let us consider two types of geminals in terms of 
their spatial broadenings. The first type is (a) geminal which is spatially 
extended with respect to both ${\rm {\bf R}}$ and ${\boldsymbol \rho}$. 
In this case, the magnitude of the geminal can be estimated as 
\begin{equation}
\label{eq7}
\nu_{\Psi } \left( {{\rm {\bf R}}{\boldsymbol \rho};\zeta_{1} \zeta_{2} } \right)\sim 
{\frac{1}{\sqrt \Omega }}{\frac{1}{\sqrt \Omega }},
\end{equation}
where $\Omega $ is the volume of the system. The second type is (b) geminal 
which is extended with respect to ${\rm {\bf R}}$, but is localized with 
respect to ${\boldsymbol \rho}$ in some region $\omega \,(\ll \,\Omega )$. In 
this case, the magnitude of the geminal can be estimated as
\begin{equation}
\label{eq8}
\nu_{\Psi } \left( {{\rm {\bf R}}{\boldsymbol \rho};\zeta_{1} \zeta_{2} } \right)\sim 
{\frac{1}{\sqrt \Omega }}{\frac{1}{\sqrt \omega }}.
\end{equation}

In case (a), there would not exist an attractive force between electrons, 
and any kinds of electron pairings do not form in the system. That is to 
say, the system is not under the superconducting state but under the normal 
state. On the other hand, in case (b), electrons would attract each other 
via some attractive force that overcomes the Coulomb repulsive force, which 
yields the pairings of electrons. 

The BEC of the fermion system is defined by means of the spectrum 
decomposition of the RDM2 \cite{1}. If the eigenvalue for some geminal becomes 
$O(N)$, namely, if the occupation number of some geminal becomes $O(N)$, 
then it is said that the BEC occurs in the fermion system \cite{1,2}. If such an 
eigenvalue and corresponding geminal are denoted as $n_{\nu_{\Psi }^{\max 
} }^{(2)} $ and $\left| {\nu^{\max}_{\Psi}} \right\rangle$, respectively, the 
spectrum decomposition (\ref{eq6}) is formally rewritten as

\begin{eqnarray}
\label{eq9}
 \left\langle {\hat{{D}}^{(2)}\left( {{\rm {\bf r}}_{1} \zeta_{1} ,{\rm 
{\bf r}}_{2} \zeta_{2} ;{\rm {\bf {r}'}}_{1} {\zeta }'_{1} ,{\rm {\bf 
{r}'}}_{2} {\zeta }'_{2} } \right)} \right\rangle_{\Psi } 
&=& 
n_{\nu_{\Psi }^{\max } }^{(2)} \nu_{\Psi }^{\max } \left( {{\rm {\bf R}}{\boldsymbol \rho};\zeta_{1} 
\zeta_{2} } \right)\nu_{\Psi }^{\max } \left( {\rm {\bf {R}'}{\boldsymbol \rho}'};
{\zeta }'_{1} {\zeta }'_{2}  \right)^{\ast } \nonumber \\ 
&+&
\sum\limits_{\nu_{\Psi } \ne \nu_{\Psi }^{\max } } {n_{\nu_{\Psi } 
}^{(2)} } \nu_{\Psi } \left( {{\rm {\bf R }}{\boldsymbol \rho};\zeta_{1} \zeta_{2} } 
\right)\nu_{\Psi } \left( {\rm {\bf {R}'}{\boldsymbol \rho}';{\zeta }'_{1} {\zeta 
}'_{2} } \right)^{\ast }. 
\end{eqnarray}
Let us assume that the BEC originates from only one kind of geminal, i.e., 
$\left| {\nu^{\max}_{\Psi}} \right\rangle $ \cite{62}. Then, the eigenvalues of any 
geminals except $\left| {\nu^{\max}_{\Psi}} \right\rangle$ are much less than 
$O(N)$, and the spatial broadenings of these geminals, i.e., $\nu_{\Psi } 
\left( {{\rm {\bf r}}_{1} \zeta_{1} ,{\rm {\bf r}}_{2} \zeta_{2} } 
\right)$'s that appear in the second term of Eq. (\ref{eq9}), are necessarily type 
(a). This is because if some geminal other than 
$\left| {\nu^{\max}_{\Psi}} \right\rangle $ 
belongs to type (b) the BEC associated with it would appear 
simultaneously with the BEC of $\left| {\nu^{\max}_{\Psi}} \right\rangle $ \cite{63}, 
which contradicts the above assumption. 

Here we shall give the definition of the superconducting state. When the BEC 
occurs with some geminal, and further when such a geminal is type (b), then 
we will say that the superconducting state appears in the system. This 
definition sounds reasonable, because the number of pairings is $O(N)$ and 
they are occupied in the same two-particle state, which is analogous with 
the intuitive description of the BEC of the ideal bosons, and further 
because such two-particle state is spatially localized like the Cooper pair 
\cite{64,65}. Thus, in the superconducting state, we have
\begin{equation}
\label{eq10}
\left\{ {{\begin{array}{*{20}c}
 {n_{\nu_{\Psi }^{\max } }^{(2)} =O(N),\,\,\,\,\,\nu_{\Psi }^{\max } 
\left( {{\rm {\bf r}}_{1} \zeta_{1} ,{\rm {\bf r}}_{2} \zeta_{2} } 
\right)=\mbox{type}\,\,\mbox{(b),}} \hfill \\
 {n_{\nu_{\Psi } }^{(2)} \ll O(N),\,\,\,\,\,\nu_{\Psi } \left( {{\rm {\bf 
r}}_{1} \zeta_{1} ,{\rm {\bf r}}_{2} \zeta_{2} } 
\right)=\mbox{type}\,\,\mbox{(a),}} \hfill \\
\end{array} }} \right.
\end{equation}
The striking point of the present definition is to define the spatial 
broadening of geminals which are condensed in the superconducting state. 
%
\subsection{Order parameter of the superconducting state}
\label{secII-C}
In this subsection, according to the definition of the superconducting 
state, we shall revisit the OPSS. As mentioned in Sec. II-B, the eigenvalue 
of some geminal becomes $O(N)$ when the superconducting state appears in 
the system. 

First, let us take the following limit on both sides of Eq. (\ref{eq9}) \cite{1}: 
\begin{equation}
\label{eq11}
{\rm {\bf r}}_{1} \approx {\rm {\bf r}}_{2} ,\,\,\,{\rm {\bf {r}'}}_{1} 
\approx {\rm {\bf {r}'}}_{2} ,\,\,\,\,\mbox{and}\,\,\,\,\left| {{\rm {\bf 
r}}_{1} -{\rm {\bf {r}'}}_{1} } \right|\longrightarrow \infty .
\end{equation}
This limit means that the distance between particles belonging to the same 
geminal gets close to each other and the distance between geminals spreads 
infinitely. It should be noted that the limit (\ref{eq11}) includes the 
thermodynamical limit given by 
\begin{equation}
\label{eq12}
N\to \infty ,\,\,\,\Omega \to \infty 
,\,\,\,\mbox{with}\,\,\,\,{\frac{N}{\Omega }}=n\,\,\mbox{(constant)},
\end{equation}
where $N$ and $\Omega $ are particle number and volume of the system, 
respectively. In the superconducting state, the first term of the right-hand 
side of Eq. (\ref{eq9}) remains and takes the following value when taking the limit 
of Eq. (\ref{eq11}):
\begin{equation}
\label{eq13}
n_{\nu_{\Psi }^{\max } }^{(2)} \nu_{\Psi }^{\max } \left( {{\rm {\bf R}}{\boldsymbol \rho};
\zeta_{1} \zeta_{2} } \right)\nu_{\Psi }^{\max } \left( {\rm {\bf {R}'}{\boldsymbol \rho}';{\zeta }'_{1} {\zeta }'_{2} } \right)^{\ast }\,\,\approx 
\,\,O(N)\left( {{\frac{1}{\sqrt \Omega }}{\frac{1}{\sqrt \omega }}} 
\right)^{2},
\end{equation}
where Eqs. (\ref{eq10}) and (\ref{eq8}) are used. Equation (\ref{eq13}) takes the value of the order 
of $n/\omega$, which is a finite and nonzero value. On 
the other hand, the second term of Eq. (\ref{eq9}) is shown to be zero in the limit 
of Eq. (\ref{eq11}) when the system is under the superconducting state. It is proved 
as follows.

Since we merely intend to check whether the second term of Eq. (\ref{eq9}) vanish or 
not in the limit of Eq. (\ref{eq11}), it is sufficient to estimate the order of the 
magnitude of $\nu_{\Psi } \left( {{\rm {\bf R}}{\boldsymbol \rho};\zeta_{1} \zeta_{2} 
} \right)$. For this aim, it seems to be appropriate to adopt the Slater 
determinant as an explicit form of 
$\nu_{\Psi } \left( {\rm {\bf R}{\boldsymbol \rho};\zeta_{1} \zeta_{2} } \right)$. 
In other words, the Hartree-Fock 
approximation may be used for the order estimation of the magnitude of $\nu 
_{\Psi } \left( {{\rm {\bf R}}{\boldsymbol \rho};\zeta_{1} \zeta_{2} } \right)$. A 
similar estimation using the Hartree-Fock approximation has been done by 
Yang in the paper where the BEC of the fermion system was first discussed 
\cite{1}. Since $\nu_{\Psi } \left( {\rm {\bf R}{\boldsymbol \rho};\zeta_{1} \zeta_{2} } 
\right)$ is spatially extended with respect to both ${\rm {\bf R}}$ and 
${\boldsymbol \rho}$, as mentioned in Eq. (\ref{eq10}), it is appropriate to suppose 
that the Slater determinant is constructed from the plane waves. Namely we 
have
\begin{eqnarray}
\label{eq14}
\nu_{\Psi } \left( {{\rm {\bf R}}{\boldsymbol \rho};\zeta_{1} \zeta_{2} } \right)
&=&
\nu_{\Psi } \left( {{\rm {\bf r}}_{1} \zeta_{1} ,{\rm {\bf r}}_{2} \zeta_{2} } \right) 
\nonumber \\ 
&=&
{\frac{1}{\sqrt 2 }}
\left| {{\begin{array}{*{20}c}
 {\phi_{a} ({\rm {\bf r}}_{1} \zeta_{1} )} \hfill & {\phi_{b} ({\rm {\bf r}}_{1} \zeta_{1} )} \hfill \\
 {\phi_{a} ({\rm {\bf r}}_{2} \zeta_{2} )} \hfill & {\phi_{b} ({\rm {\bf r}}_{2} \zeta_{2} )} \hfill \\
\end{array} }} \right|,  
\end{eqnarray}
where $\phi_{a} ({\rm {\bf r}}\zeta )$ is the plane wave containing the 
spin function $\chi_{\sigma_{a} } (\zeta )$, which is given by
\begin{equation}
\label{eq15}
\phi_{a} ({\rm {\bf r}}\zeta )={\frac{1}{\sqrt \Omega }}e^{i{\rm {\bf 
k}}_{a} \cdot {\rm {\bf r}}}\chi_{\sigma_{a} } (\zeta ).
\end{equation}
Substituting Eq. (\ref{eq15}) into Eq. (\ref{eq14}), we get
\begin{equation}
\label{eq16}
\nu_{\Psi } \left( {{\rm {\bf R}}{\boldsymbol \rho};\zeta_{1} \zeta_{2} } 
\right)={\frac{1}{\sqrt 2 \Omega }}e^{i({\rm {\bf k}}_{a} +{\rm {\bf k}}_{b} 
)\cdot {\rm {\bf R}}}\left\{ {e^{i({\rm {\bf k}}_{a} -{\rm {\bf k}}_{b} 
)\cdot {\frac{{\boldsymbol \rho}}{2}}}\chi_{\sigma_{a} } (\zeta_{1} )\chi 
_{\sigma_{b} } (\zeta_{2} )-e^{-i({\rm {\bf k}}_{a} -{\rm {\bf k}}_{b} 
)\cdot {\frac{{\boldsymbol \rho}}{2}}}\chi_{\sigma_{a} } (\zeta_{2} )\chi 
_{\sigma_{b} } (\zeta_{1} )} \right\}.
\end{equation}
Note that Eq. (\ref{eq16}) is the spatially-extended function with respect to both 
${\rm {\bf R}}$ and ${\boldsymbol \rho}$.

Using Eq. (\ref{eq16}), the second term of Eq. (\ref{eq9}) is expressed as
\begin{eqnarray}
\label{eq17}
&&\sum\limits_{\nu_{\Psi } \ne \nu_{\Psi }^{\max } } 
{n_{\nu_{\Psi } }^{(2)} \nu_{\Psi } \left( {\rm {\bf R}}{\boldsymbol \rho};\zeta_{1} \zeta_{2} \right)} \nu_{\Psi } \left( \rm {\bf {R}'}{\boldsymbol \rho}';{\zeta }'_{1} {\zeta}'_{2}  \right)^{\ast } \nonumber \\ 
&=&\sum\limits_{{\rm {\bf k}}_{a} ,{\rm {\bf k}}_{b},\sigma_{a} ,\sigma_{b}} 
{\frac{n_{{\rm {\bf k}}_{a} \sigma_{a} {\rm {\bf k}}_{b} \sigma_{b} }^{(2)}}{2\Omega }} 
\left\{ 
{e^{i{\rm {\bf k}}_{a} \cdot ({\rm {\bf r}}_{1} -{\rm {\bf {r}'}}_{1} )}
 e^{i{\rm {\bf k}}_{b} \cdot ({\rm {\bf r}}_{2} -{\rm {\bf {r}'}}_{2} )}
\chi_{\sigma_{a} } (\zeta_{1} )\chi _{\sigma_{b} } (\zeta_{2} )
\chi_{\sigma_{a} } ({\zeta }'_{1} )\chi _{\sigma_{b} } ({\zeta }'_{2} )} \right. \nonumber \\ 
&&\,\,\,\,\,\,\,\,\,\,\,\,\,\,\,\,\,\,\,\,\,\,\,\,\,\,\,\,\,\,\,\,\,\,\,\,\,\,\,\,\,\,\,\,\,\,\,
-e^{i{\rm {\bf k}}_{a} \cdot ({\rm {\bf r}}_{1} -{\rm {\bf {r}'}}_{2} )}
 e^{i{\rm {\bf k}}_{b} \cdot ({\rm {\bf r}}_{2} -{\rm {\bf {r}'}}_{1} )}
\chi_{\sigma_{a} } (\zeta_{1} )\chi_{\sigma_{b} } (\zeta_{2} )
\chi_{\sigma_{a} } ({\zeta }'_{2} )\chi_{\sigma_{b} } ({\zeta }'_{1} )    \\ 
&&\,\,\,\,\,\,\,\,\,\,\,\,\,\,\,\,\,\,\,\,\,\,\,\,\,\,\,\,\,\,\,\,\,\,\,\,\,\,\,\,\,\,\,\,\,\,\,
-e^{i{\rm {\bf k}}_{a} \cdot ({\rm {\bf r}}_{2} -{\rm {\bf {r}'}}_{1} )}
 e^{i{\rm {\bf k}}_{b} \cdot ({\rm {\bf r}}_{1} -{\rm {\bf {r}'}}_{2} )}
\chi_{\sigma_{a} } (\zeta_{2} )\chi_{\sigma_{b} } (\zeta_{1} )
\chi_{\sigma_{a} } ({\zeta }'_{1} )\chi_{\sigma_{b} } ({\zeta }'_{2} ) \nonumber \\ 
&&\,\,\,\,\,\,\,\,\,\,\,\,\,\,\,\,\,\,\,\,\,\,\,\,\,\,\,\,\,\,\,\,\,\,\,\,\,\,\,\,\,\,\,\,\,\,\,\left. 
+e^{i{\rm {\bf k}}_{a} \cdot ({\rm {\bf r}}_{2} -{\rm {\bf {r}'}}_{2} )}
e^{i{\rm {\bf k}}_{b} \cdot ({\rm {\bf r}}_{1} -{\rm {\bf {r}'}}_{1} )}
\chi_{\sigma_{a} } (\zeta_{2} )\chi_{\sigma_{b} } (\zeta_{1} )
\chi _{\sigma_{a} } ({\zeta }'_{2} )\chi_{\sigma_{b} } ({\zeta }'_{1} ) 
\right\}.  \nonumber
\end{eqnarray}
We shall take the limit of Eq. (\ref{eq11}) on the right-hand side of Eq. (\ref{eq17}). 
Using the Riemann-Lebesgue theorem for the oscillating function \cite{66}, all 
four terms are shown to vanish in this limit. Namely, 
\begin{equation}
\label{eq18}
\sum\limits_{\nu_{\Psi } \ne \nu_{\Psi }^{\max } } {n_{\nu_{\Psi } 
}^{(2)} } \nu_{\Psi } \left( {{\rm {\bf R}}{\boldsymbol \rho};\zeta_{1} \zeta_{2} } 
\right)\nu_{\Psi } \left( {{\rm {\bf {R}'}}{\boldsymbol \rho}';{\zeta }'_{1} {\zeta 
}'_{2} } \right)^{\ast }\,\,\,\mathop {\longrightarrow}\limits_{\rm{Eq}. (11)} 
\,\,\,\,\,0.
\end{equation}
Considering Eqs. (\ref{eq13}) and (\ref{eq18}) together, the RDM2 $\left\langle 
{\hat{{D}}^{(2)}\left( {{\rm {\bf r}}_{1} \zeta_{1} ,{\rm {\bf r}}_{2} 
\zeta_{2} ;{\rm {\bf {r}'}}_{1} {\zeta }'_{1} ,{\rm {\bf {r}'}}_{2} {\zeta 
}'_{2} } \right)} \right\rangle_{\Psi } $ can take a finite and nonzero 
value in the limit of Eq. (\ref{eq11}) when the system ${\left| {\Psi} \right\rangle }$ 
is in the superconducting state. This statement is written by the following 
formula:
\begin{equation}
\label{eq19}
\left\langle {\hat{{D}}^{(2)}\left( {{\rm {\bf r}}_{1} \zeta_{1} ,{\rm {\bf 
r}}_{2} \zeta_{2} ;{\rm {\bf {r}'}}_{1} {\zeta }'_{1} ,{\rm {\bf {r}'}}_{2} 
{\zeta }'_{2} } \right)} \right\rangle_{\Psi } \,\,\mathop {\longrightarrow}\limits_{\rm{Eq}.(11)} 
\,\,n_{\nu_{\Psi }^{\max } }^{(2)} \nu_{\Psi }^{\max 
} \left( {{\rm {\bf R}}{\boldsymbol \rho};\zeta_{1} \zeta_{2} } \right)\nu_{\Psi 
}^{\max } \left( {{\rm {\bf {R}'}}{\boldsymbol \rho}';{\zeta }'_{1} {\zeta }'_{2} } 
\right)^{\ast }\,.
\end{equation}
Of course, the right-hand side of Eq. (\ref{eq19}) is nonzero, and is estimated to 
be the order of $n/\omega$ in the superconducting state, as 
mentioned above. Equation (\ref{eq19}) means that 
off-diagonal elements of the RDM2 of the superconducting state becomes the order of 
$n/\omega$ in the limit of Eq. (\ref{eq11}). 

On the other hand, let us consider the case of taking the limit of Eq. (\ref{eq11}) 
directly on both sides of Eq. (\ref{eq2}). Generally, there does not exist any 
correlation between two systems which separate from each other in an 
infinite distance \cite{67}. If $\hat{{A}}({\rm {\bf r}})$ and $\hat{{B}}({\rm 
{\bf {r}'}})$ are arbitrary operators depending on the positions ${\rm {\bf 
r}}$ and ${\rm {\bf {r}'}}$, respectively, and further if the distance 
between ${\rm {\bf r}}$ and ${\rm {\bf {r}'}}$ is infinite, then the 
expectation value of the product of them is equal to the product of the 
individual expectation values:
\begin{equation}
\label{eq20}
{\left\langle {\Psi } \right|}\hat{{A}}({\rm {\bf r}})\hat{{B}}({\rm {\bf 
{r}'}}){\left| {\Psi} \right\rangle }\mathop {\longrightarrow}\limits_{\left| {{\rm {\bf 
r}}-{\rm {\bf {r}'}}} \right|\to \infty } {\left\langle {\Psi } 
\right|}\hat{{A}}({\rm {\bf r}}){\left| {\Psi} \right\rangle }{\left\langle 
{\Psi } \right|}\hat{{B}}({\rm {\bf {r}'}}){\left| {\Psi} \right\rangle }.
\end{equation}
Equation (\ref{eq20}) is sometimes called the cluster decomposition principle \cite{67}. 
When taking the limit of Eq. (\ref{eq11}), we can investigate the behavior of the 
RDM2 in the case where the correlation between geminals does not exist at 
all. Using the general principle (\ref{eq20}), Eq. (\ref{eq2}) becomes
\begin{equation}
\label{eq21}
\left\langle {\hat{{D}}^{(2)}\left( {{\rm {\bf r}}_{1} \zeta_{1} ,{\rm {\bf 
r}}_{2} \zeta_{2} ;{\rm {\bf {r}'}}_{1} {\zeta }'_{1} ,{\rm {\bf {r}'}}_{2} 
{\zeta }'_{2} } \right)} \right\rangle_{\Psi } \,\,\mathop {\longrightarrow}\limits_{\rm{Eq}. (11)} 
\,\,{\frac{1}{2}}{\left\langle {\Psi } \right|}\psi 
^{\dag }({\rm {\bf {r}'}}_{1} {\zeta }'_{1} )\psi^{\dag }({\rm {\bf 
{r}'}}_{2} {\zeta }'_{2} ){\left| {\Psi} \right\rangle }{\left\langle {\Psi } 
\right|}\psi ({\rm {\bf r}}_{2} \zeta_{2} )\psi ({\rm {\bf r}}_{1} \zeta 
_{1} ){\left| {\Psi} \right\rangle }.
\end{equation}
Compared Eq. (\ref{eq21}) with Eq. (\ref{eq19}), we immediately understand that the OPSS is 
given by ${\left\langle {\Psi } \right|}\psi ({\rm {\bf r}}\zeta )\psi ({\rm 
{\bf {r}'}}{\zeta }'){\left| {\Psi} \right\rangle }$. Namely we get
\begin{equation}
\label{eq22}
{\left\langle {\Psi } \right|}\psi ({\rm {\bf r}}\zeta )\psi ({\rm {\bf 
{r}'}}{\zeta }'){\left| {\Psi} \right\rangle }=\sqrt {2n_{\nu_{\Psi }^{\max } 
}^{(2)} } \nu_{\Psi }^{\max } \left( {{\rm {\bf r}}\zeta ,{\rm {\bf 
{r}'}}{\zeta }'} \right).
\end{equation}
If ${\left\langle {\Psi } \right|}\psi ({\rm {\bf r}}\zeta )\psi ({\rm {\bf 
{r}'}}{\zeta }'){\left| {\Psi} \right\rangle }$ takes a nonzero value, then the 
system ${\left| {\Psi} \right\rangle }$ is in the superconducting state. 

Further from Eq. (\ref{eq22}), the spatial and spin symmetries for the pairing 
states can be confirmed by means of the OPSS. That is to say, the OPSS 
${\left\langle {\Psi } \right|}\psi ({\rm {\bf r}}\zeta )\psi ({\rm {\bf 
{r}'}}{\zeta }'){\left| {\Psi} \right\rangle }$ is formally rewritten as a 
function of ${\rm {\bf R}}$ and ${\boldsymbol \rho }$, which directly gives 
the spatial distribution of the pairing states via Eq. (\ref{eq22}). Also concerning 
the spin symmetry, it is easily shown that the OPSS ${\left\langle {\Psi } 
\right|}\psi ({\rm {\bf r}}\zeta )\psi ({\rm {\bf {r}'}}{\zeta }'){\left| {\Psi} 
\right\rangle }$ is decomposed into the parts of spin-singlet and 
spin-triplet wave functions. 
%
\subsection{The case of the finite temperature}
\label{secII-D}
Let us extend the above-mentioned discussions to the case of the finite 
temperature. The superconducting state should be treated on the basis of the 
grand canonical ensemble because the number of electrons that form the 
superconducting phase varies depending on the condition of the system. If 
the statistical density matrix of the grand canonical ensemble is denoted as 
$\hat{{\rho }}_{H} $, the statistical average of the physical quantity 
$\hat{{A}}$ is given by $\mbox{Tr}\left( {\hat{{\rho }}_{H} \hat{{A}}} \right)$, 
which is hereafter denoted as $\left\langle {\hat{{A}}} \right\rangle_{H} 
$. The statistical average of the RDM2 with respect to $\hat{{\rho }}_{H} $ 
is given by
\begin{equation}
\label{eq23}
\left\langle {\hat{{D}}^{(2)}\left( {{\rm {\bf r}}_{1} \zeta_{1} ,{\rm {\bf 
r}}_{2} \zeta_{2} ;{\rm {\bf {r}'}}_{1} {\zeta }'_{1} ,{\rm {\bf {r}'}}_{2} 
{\zeta }'_{2} } \right)} \right\rangle_{H} =\mbox{Tr}\left\{ {\hat{{\rho }}_{H} 
\hat{{D}}^{(2)}\left( {{\rm {\bf r}}_{1} \zeta_{1} ,{\rm {\bf r}}_{2} \zeta 
_{2} ;{\rm {\bf {r}'}}_{1} {\zeta }'_{1} ,{\rm {\bf {r}'}}_{2} {\zeta }'_{2} 
} \right)} \right\}.
\end{equation}
In a similar way to the case of zero-temperature (Eq. (\ref{eq3})), we can define 
the operator $\hat{{D}}_{H}^{(2)} $ with the use of the equation 
${\left\langle {{\rm {\bf r}}_{1} \zeta_{1} {\rm {\bf r}}_{2} \zeta_{2} } 
\right|}\hat{{D}}_{H}^{(2)} 
{\left| {{\rm {\bf r}'}_{1} {\zeta}'_{1} {\rm {\bf r}}'_{2} {\zeta}'_{2} } \right\rangle }
=2\left\langle {\hat{{D}}^{(2)}\left( {{\rm {\bf r}}_{1} \zeta_{1} ,{\rm {\bf 
r}}_{2} \zeta_{2} ;{\rm {\bf {r}'}}_{1} {\zeta }'_{1} ,{\rm {\bf {r}'}}_{2} 
{\zeta }'_{2} } \right)} \right\rangle_{H} $. 
The eigenvalue equation for $\hat{{D}}_{H}^{(2)} $ is formally written as
\begin{equation}
\label{eq24}
\hat{{D}}_{H}^{(2)} {\left| {\nu_{H}} \right\rangle }
=n_{\nu_{H}}^{(2)}\left| {\nu_{H}} \right\rangle,
\end{equation}
where $\left| {\nu_{H}} \right\rangle$ and $n_{\nu_{H} }^{(2)} $ are 
eigenfunctions and eigenvalues for $\hat{{D}}_{H}^{(2)} $, respectively. 
Using Eq. (\ref{eq24}), the spectrum decomposition for $\left\langle 
{\hat{{D}}^{(2)}\left( {{\rm {\bf r}}_{1} \zeta_{1} ,{\rm {\bf r}}_{2} 
\zeta_{2} ;{\rm {\bf {r}'}}_{1} {\zeta }'_{1} ,{\rm {\bf {r}'}}_{2} {\zeta 
}'_{2} } \right)} \right\rangle_{H} $ is given by
\begin{equation}
\label{eq25}
\left\langle {\hat{{D}}^{(2)}\left( {{\rm {\bf r}}_{1} \zeta_{1} ,{\rm {\bf 
r}}_{2} \zeta_{2} ;{\rm {\bf {r}'}}_{1} {\zeta }'_{1} ,{\rm {\bf {r}'}}_{2} 
{\zeta }'_{2} } \right)} \right\rangle_{H} =\sum\limits_{\nu_{H} } {n_{\nu 
_{H} }^{(2)} } \nu_{H} \left( {{\rm {\bf r}}_{1} \zeta_{1} ,{\rm {\bf 
r}}_{2} \zeta_{2} } \right)\nu_{H} \left( {{\rm {\bf {r}'}}_{1} {\zeta 
}'_{1} ,{\rm {\bf {r}'}}_{2} {\zeta }'_{2}} \right)^{\ast },
\end{equation}
in a similar way to Eq. (\ref{eq6}). 

The superconducting state can be defined in the same way as the case of 
zero-temperature (Sec. II-B). Specifically, the definition of the 
superconducting state is given by the following replacement in Eq. (\ref{eq10}):
\begin{equation}
\label{eq26}
\left\{ {{\begin{array}{*{20}c}
 {n_{\nu_{\Psi } }^{(2)} } \hfill \\
 {n_{\nu_{\Psi }^{\max } }^{(2)} } \hfill \\
 {\nu_{\Psi } \left( {{\rm {\bf r}}_{1} \zeta_{1} ,{\rm {\bf r}}_{2} \zeta 
_{2} } \right)} \hfill \\
 {\,\,\nu_{\Psi }^{\max } \left( {{\rm {\bf r}}_{1} \zeta_{1} ,{\rm {\bf 
r}}_{2} \zeta_{2} } \right)} \hfill \\
\end{array} }} \right.\,\,\,\,\,\longrightarrow \,\,\,\,\,\left\{ 
{{\begin{array}{*{20}c}
 {n_{\nu_{H} }^{(2)} } \hfill \\
 {n_{\nu_{H}^{\max } }^{(2)} } \hfill \\
 {\nu_{H} \left( {{\rm {\bf r}}_{1} \zeta_{1} ,{\rm {\bf r}}_{2} \zeta 
_{2} } \right)} \hfill \\
 {\,\,\nu_{H}^{\max } \left( {{\rm {\bf r}}_{1} \zeta_{1} ,{\rm {\bf 
r}}_{2} \zeta_{2} } \right)}. \hfill \\
\end{array} }} \right.
\end{equation}
Also concerning the OPSS, we can obtain it by using Eq. (\ref{eq26}) and by 
replacing the expectation value ${\left\langle {\Psi } 
\right|}\hat{{A}}{\left| {\Psi} \right\rangle }$ with the statistical average 
$\left\langle {\hat{{A}}} \right\rangle_{H} $ in the discussions of Sec. 
II-C. The explicit form of the OPSS at the finite temperature is given by
\begin{equation}
\label{eq27}
\left\langle {\psi ({\rm {\bf r}}\zeta )\psi ({\rm {\bf {r}'}}{\zeta }')} 
\right\rangle_{H} =\mbox{Tr}\left\{ {\hat{{\rho }}_{H} \psi ({\rm {\bf 
r}}\zeta )\psi ({\rm {\bf {r}'}}{\zeta }')} \right\}.
\end{equation}
Here we suppose that the cluster decomposition principle holds also at the 
finite temperature if the system is stable \cite{67}:
\begin{equation}
\label{eq28}
\left\langle {\hat{{A}}({\rm {\bf r}})\hat{{B}}({\rm {\bf {r}'}})} 
\right\rangle_{\!\!H} \,\,\
\mathop {\longrightarrow}\limits_{\left| {{\rm {\bf r}}-{\rm {\bf {r}'}}} \right| {\to} \infty } \,\,\,
\left\langle {\hat{{A}}({\rm {\bf r}})} 
\right\rangle_{\!\!H} \left\langle {\hat{{B}}({\rm {\bf {r}'}})} \right\rangle 
_{\!\!H} ,
\end{equation}
where $\hat{{A}}({\rm {\bf r}})$ and $\hat{{B}}({\rm {\bf r}'})$ are 
operators of arbitrary physical quantities. In a similar way to Eq. (\ref{eq22}), 
the OPSS (\ref{eq27}) is rewritten as
\begin{equation}
\label{eq29}
\left\langle {\psi ({\rm {\bf r}}\zeta )\psi ({\rm {\bf {r}'}}{\zeta }')} 
\right\rangle_{H} =\sqrt {2n_{\nu_{H}^{\max } }^{(2)} } \nu_{H}^{\max } 
\left( {{\rm {\bf r}}\zeta ,{\rm {\bf {r}'}}{\zeta }'} \right).
\end{equation}
Thus, the extension to the finite temperature can be done by the replacement
\begin{equation}
\label{eq30}
{\left\langle {\Psi } \right|}\psi ({\rm {\bf r}}\zeta )\psi ({\rm {\bf 
{r}'}}{\zeta }'){\left| {\Psi} \right\rangle }\,\,\,\,\,\longrightarrow 
\,\,\,\,\,\,\left\langle {\psi ({\rm {\bf r}}\zeta )\psi ({\rm {\bf 
{r}'}}{\zeta }')} \right\rangle_{H} ,
\end{equation}
in addition to Eq. (\ref{eq26}).
%
\section{ Fluctuations of the particle number in the superconducting state}
\label{secIII}
In this section, it is shown that diagonal elements of the RDM2 can be used 
as an indication quantity of the superconducting state instead of the 
conventionally-used OPSS, i.e., off-diagonal elements of the RDM2. 
Specifically, we shall show that the fluctuation of the particle number 
becomes $O(N)$ when the OPSS appears in the system.

%
\subsection{Bloch-de Dominicis theorem for the RDM2}
\label{secIII-A}
It is sufficient for the order estimation of the physical quantities to make 
a use of the Hartree-Fock approximation. In Sec.II-C, we have already 
adopted the Hartree-Fock approximation for estimating the order of the 
second term of Eq. (\ref{eq9}). In this subsection we shall consider the 
Hartree-Fock approximation of the RDM2 at finite temperature in the 
superconducting state.

The Hartree-Fock approximation of some physical quantity at the finite 
temperature corresponds to the statistical average of the quantity by means 
of the density matrix involving the mean-field Hamiltonian. In the 
superconducting state, the mean-field Hamiltonian is devised such that the 
gauge symmetry is explicitly broken, as can be seen in the BCS theory \cite{65}, 
Bogoliubov-de Gennes theory \cite{7,8}, and the density-functional schemes 
\cite{9,10,25}. If such a mean-field Hamiltonian is denoted as $\hat{{H}}_{M} $, 
then it does not preserve the electron number, but can be diagonalized in 
terms of the quasiparticles via the Bogoliubov-Valatin transformation 
\cite{68,69}. The Bogoliubov-Valatin transformation from the electron system to 
the quasiparticle system is generally written as \cite{68,69}
\begin{equation}
\label{eq31}
\begin{array}{l}
 \psi ({\rm {\bf r}}\zeta )=\sum\limits_i {u_{i} ({\rm {\bf r}}\zeta )\gamma 
_{i} +\sum\limits_j {v_{j} ({\rm {\bf r}}\zeta )\gamma_{j}^{\dag } } } , \\ 
 \psi^{\dag }({\rm {\bf r}}\zeta )=\sum\limits_i {u_{i}^{\ast }({\rm {\bf 
r}}\zeta )\gamma_{i}^{\dag } +\sum\limits_j {v_{j}^{\ast }({\rm {\bf 
r}}\zeta )\gamma_{j} } } , \\ 
 \end{array}
\end{equation}
where $\gamma_{i} $ and $\gamma_{i}^{\dag } $ are annihilation and 
creation operators of quasiparticles, and where $u_{i} ({\rm {\bf r}}\zeta 
)$ and $v_{i} ({\rm {\bf r}}\zeta )$ are elements of the unitary matrix 
which are determined by requiring that $\hat{{H}}_{M} $ is diagonalized in 
terms of the quasiparticles \cite{8}. Suppose that $\hat{{H}}_{M} $ is written as
\begin{equation}
\label{eq32}
\hat{{H}}_{M} -\mu \hat{{N}}=\sum\limits_i {\xi_{i} \gamma_{i}^{\dag } 
\gamma_{i} } ,
\end{equation}
where $\hat{{N}}$ is the operator of the particle number, and where $\mu $ 
is the chemical potential of the system. By using this $\hat{{H}}_{M} $, we 
consider the statistical average of the RDM2, i.e., the Hartree-Fock 
approximation of the RDM2:
\begin{equation}
\label{eq33}
\left\langle {\hat{{D}}^{(2)}\left( {{\rm {\bf r}}_{4} \zeta_{4} ,{\rm {\bf 
r}}_{3} \zeta_{3} ;{\rm {\bf r}}_{1} \zeta_{1} ,{\rm {\bf r}}_{2} \zeta 
_{2} } \right)} \right\rangle_{\!H_{M} } ={\frac{1}{2}}\mbox{Tr}\left\{ 
{\hat{{\rho }}_{H_{M} } \psi^{\dag }({\rm {\bf r}}_{1} \zeta_{1} )\psi 
^{\dag }({\rm {\bf r}}_{2} \zeta_{2} )\psi ({\rm {\bf r}}_{3} \zeta_{3} 
)\psi ({\rm {\bf r}}_{4} \zeta_{4} )} \right\},
\end{equation}
where the concrete forms of the statistical density matrix 
$\hat{{\rho }}_{H_{M} }$ is given by
\begin{equation}
\label{eq34}
\hat{{\rho }}_{H_{M} } 
={\frac{1}{\Xi }}e^{-\beta \left( {\hat{H}_{M} -\mu \hat{{N}}} \right)},
\end{equation}
with $\Xi =\mbox{Tr}\left( {e^{-\beta \left( {\hat{H}_{M} -\mu \hat{{N}}} \right)}} 
\right)$.  
Using Eqs. (\ref{eq31}), (\ref{eq32}) and (\ref{eq34}), it is shown that the following cluster 
decomposition of the RDM2, which is so-called the Bloch-de Dominicis theorem 
\cite{70,71,72}, holds:
\begin{eqnarray}
\label{eq35}
\!\!\!
\left\langle {\psi^{\dag }({\rm {\bf r}}_{1} \zeta_{1} )\psi^{\dag 
}({\rm {\bf r}}_{2} \zeta_{2} )\psi ({\rm {\bf r}}_{3} \zeta_{3} )\psi 
({\rm {\bf r}}_{4} \zeta_{4} )} \right\rangle_{\!H_{M} } 
&=&
\left\langle {\psi^{\dag }({\rm {\bf r}}_{1} \zeta_{1} )\psi ({\rm {\bf r}}_{4} \zeta_{4} )} 
\right\rangle_{\!H_{M} } \left\langle {\psi^{\dag }({\rm {\bf r}}_{2} \zeta 
_{2} )\psi ({\rm {\bf r}}_{3} \zeta_{3} )} \right\rangle_{\!H_{M} } \nonumber \\ 
&-&\left\langle {\psi^{\dag }({\rm {\bf r}}_{1} \zeta_{1} )\psi ({\rm {\bf 
r}}_{3} \zeta_{3} )} \right\rangle_{\!H_{M} } \left\langle {\psi^{\dag 
}({\rm {\bf r}}_{2} \zeta_{2} )\psi ({\rm {\bf r}}_{4} \zeta_{4} )} 
\right\rangle_{\!H_{M} } \\ 
&+&\left\langle {\psi^{\dag }({\rm {\bf r}}_{1} \zeta_{1} )\psi^{\dag 
}({\rm {\bf r}}_{2} \zeta_{2} )} \right\rangle_{\!H_{M} } \left\langle {\psi 
({\rm {\bf r}}_{3} \zeta_{3} )\psi ({\rm {\bf r}}_{4} \zeta_{4} )} 
\right\rangle_{\!H_{M} }. \nonumber
\end{eqnarray}
This equation means that the Bloch-de Dominicis theorem holds even when 
taking the statistical average of the RDM2 written by the field operators of 
electrons by means of the statistical density matrix of the noninteracting 
quasiparticle system.
%
\subsection{Idempotent of the RDM1}
\label{secIII-B}
On the right-hand side of Eq. (\ref{eq35}), we have the statistical average of the 
first-order reduced density matrix (RDM1). For the convenience of the later 
discussion, let us consider the idempotent of the RDM1. As is well known, 
the idempotent of the RDM1 rigorously holds at zero temperature within the 
Hartree-Fock approximation \cite{61}. However, it is not obvious that such the 
idempotent holds in the superconducting state at the finite temperature. To 
tell the conclusion first, the idempotent of the RDM1 approximately holds 
even in the superconducting state at low temperature. We will show it below.

The operator of the RDM1 is given by \cite{61}
\begin{equation}
\label{eq36}
\hat{{D}}^{(1)}\left( {{\rm {\bf r}}_{1} \zeta_{1} ,{\rm {\bf r}}_{2} \zeta 
_{2} } \right)=\psi^{\dag }({\rm {\bf r}}_{2} \zeta_{2} )\psi ({\rm {\bf 
r}}_{1} \zeta_{1} ).
\end{equation}
If the RDM1 of the system of state ${\left| {\Psi} \right\rangle }$ is denoted as 
$\left\langle {\hat{{D}}^{(1)}\left( {{\rm {\bf r}}_{1} \zeta_{1} ,{\rm 
{\bf r}}_{2} \zeta_{2} } \right)} \right\rangle_{\!\Psi } $, it is given by
\begin{equation}
\label{eq37}
\left\langle {\hat{{D}}^{(1)}\left( {{\rm {\bf r}}_{1} \zeta_{1} ,{\rm {\bf 
r}}_{2} \zeta_{2} } \right)} \right\rangle_{\!\Psi } ={\left\langle {\Psi } 
\right|}\hat{{D}}^{(1)}\left( {{\rm {\bf r}}_{1} \zeta_{1} ,{\rm {\bf 
r}}_{2} \zeta_{2} } \right){\left| {\Psi} \right\rangle }.
\end{equation}
Also if the statistical average of $\hat{{D}}^{(1)}\left( {{\rm {\bf r}}_{1} 
\zeta_{1} ,{\rm {\bf r}}_{2} \zeta_{2} } \right)$ with the density matrix 
$\hat{{\rho }}_{H_{M} } $ is denoted as $\left\langle {\hat{{D}}^{(1)}\left( 
{{\rm {\bf r}}_{1} \zeta_{1} ,{\rm {\bf r}}_{2} \zeta_{2} } \right)} 
\right\rangle_{\!H_{M} } $, then it is written as 
\begin{equation}
\label{eq38}
\left\langle {\hat{{D}}^{(1)}\left( {{\rm {\bf r}}_{1} \zeta_{1} ,{\rm {\bf 
r}}_{2} \zeta_{2} } \right)} \right\rangle_{H_{M} } =\mbox{Tr}\left( 
{\hat{{\rho }}_{\!H_{M} } \hat{{D}}^{(1)}\left( {{\rm {\bf r}}_{1} \zeta_{1} 
,{\rm {\bf r}}_{2} \zeta_{2} } \right)} \right).
\end{equation}
When the trace on the right-hand side is taken by means of eigenfunctions of 
$\hat{H}_{M} $, Eq. (\ref{eq38}) is rewritten as
\begin{equation}
\label{eq39}
\left\langle {\hat{{D}}^{(1)}\left( {{\rm {\bf r}}_{1} \zeta_{1} ,{\rm {\bf 
r}}_{2} \zeta_{2} } \right)} \right\rangle_{\!H_{M} } ={\frac{1}{\Xi 
}}\sum\limits_m {e^{-\beta E_{m} }\left\langle {\hat{{D}}^{(1)}\left( {{\rm 
{\bf r}}_{1} \zeta_{1} ,{\rm {\bf r}}_{2} \zeta_{2} } \right)} 
\right\rangle_{\Phi_{m} } } ,
\end{equation}
where
\begin{equation}
\label{eq40}
\hat{H}_{M} {\left| {\Phi}_{m} \right\rangle }= E_{m}{\left| {\Phi}_{m} \right\rangle }.
\end{equation}
The RDM1 of the system of state ${\left| {{\Phi}_{m}} \right\rangle }$, i.e., 
$\left\langle {\hat{{D}}^{(1)}\left( {{\rm {\bf r}}_{1} \zeta_{1} ,{\rm 
{\bf r}}_{2} \zeta_{2} } \right)} \right\rangle_{\Phi_{m} } $, 
can be rewritten by using the spectrum decomposition such that
\begin{equation}
\label{eq41}
\left\langle {\hat{{D}}^{(1)}\left( {{\rm {\bf r}}_{1} \zeta_{1} ,{\rm {\bf 
r}}_{2} \zeta_{2} } \right)} \right\rangle_{\Phi_{m} } =\sum\limits_{\mu 
_{\Phi_{m} } } {n_{\mu_{\Phi_{m} } }^{(1)} \mu_{\Phi_{m} } \! \left( {{\rm 
{\bf r}}_{1} \zeta_{1} } \right)\mu_{\Phi_{m} } \! \left( {{\rm {\bf r}}_{2} 
\zeta_{2} } \right)^{\ast }} .
\end{equation}
Substituting Eq. (\ref{eq41}) into Eq. (\ref{eq39}), we have
\begin{equation}
\label{eq42}
\left\langle {\hat{{D}}^{(1)}\left( {{\rm {\bf r}}_{1} \zeta_{1} ,{\rm {\bf 
r}}_{2} \zeta_{2} } \right)} \right\rangle_{\!H_{M} } ={\frac{1}{\Xi 
}}\sum\limits_m {e^{-\beta E_{m} }} \sum\limits_{\mu_{\Phi_{m} } } {n_{\mu 
_{\Phi_{m} } }^{(1)} \mu_{\Phi_{m} } \! \left( {{\rm {\bf r}}_{1} \zeta_{1} 
} \right)\mu_{\Phi_{m} } \! \left( {{\rm {\bf r}}_{2} \zeta_{2} } 
\right)^{\ast }}.
\end{equation}
This is the spectrum decomposition of the RDM1 at the finite temperature. 

In order to consider the idempotent of the RDM1, we calculate the following 
product of the RDM1's:
\begin{eqnarray}
\label{eq43}
&&\!\!\!\!\! \int \!{\left\langle {\hat{{D}}^{(1)}\left( {{\rm {\bf r}}_{1} \zeta_{1} 
,{\rm {\bf r}}_{2} \zeta_{2} } \right)} \right\rangle_{H_{M} } 
\left\langle {\hat{{D}}^{(1)}\left( {{\rm {\bf r}}_{2} \zeta_{2} ,{\rm {\bf 
r}}_{1} \zeta_{1} } \right)} \right\rangle_{H_{M} } d^{3}r_{1} d^{3}r_{2} 
d\zeta_{1} d\zeta_{2} } =\sum\limits_m {\sum\limits_{{m}'} 
{{\frac{e^{-\beta E_{m} }}{\Xi }}{\frac{e^{-\beta E_{{m}'} }}{\Xi }}} } \nonumber \\ 
\!\!\!&\times&  \! \! \sum\limits_{\mu_{\Phi_{m} } } 
{\sum\limits_{\mu_{\Phi_{{m}'} } } {n_{\mu_{\Phi_{m} } }^{(1)} n_{\mu 
_{\Phi_{{m}'} } }^{(1)} } } 
\!\int \!
{\mu_{\Phi_{m} } \! \left( {{\rm {\bf r}}_{1} \zeta_{1} } \right)
\mu_{\Phi_{m} } \! \left( {{\rm {\bf r}}_{2} \zeta_{2} } \right)
\mu_{\Phi_{{m}'} } \! \left( {{\rm {\bf r}}_{2} \zeta _{2} } \right)^{\ast }
\mu_{\Phi_{{m}'} } \! \left( {{\rm {\bf r}}_{1} \zeta _{1} } \right)^{\ast }d^{3}r_{1} d^{3}r_{2} d\zeta_{1} d\zeta_{2} }. 
\end{eqnarray}
On the other hand, using Eq. (\ref{eq42}), we have
\begin{equation}
\label{eq44}
\int {\left\langle {\hat{{D}}^{(1)}\left( {{\rm {\bf r}}_{1} \zeta_{1} 
,{\rm {\bf r}}_{1} \zeta_{1} } \right)} \right\rangle_{\!H_{M} } } 
d^{3}r_{1} d\zeta_{1} =\sum\limits_m {\sum\limits_{\mu_{\Phi_{m} } } 
{{\frac{e^{-\beta E_{m} }}{\Xi }}n_{\mu_{\Phi_{m} } }^{(1)} } ,} 
\end{equation}
where the normalization of the natural spin orbital is used \cite{73}. Let us 
consider the magnitude of ${e^{-\beta E_{m} }} \mathord{\left/ {\vphantom 
{{e^{-\beta E_{m} }} \Xi }} \right. \kern-\nulldelimiterspace} \Xi $ which 
is contained on the right-hand sides of both Eqs. (\ref{eq43}) and (\ref{eq44}). Suppose 
that the eigenvalues $E_{m} $ in Eq. (\ref{eq40}) take positive values, and that the 
ground state is non-degenerate. Then, the magnitude of ${e^{-\beta E_{m} }} 
\mathord{\left/ {\vphantom {{e^{-\beta E_{m} }} \Xi }} \right. 
\kern-\nulldelimiterspace} \Xi $ for the ground state is larger than those 
for the excited states, and especially at low temperature, ${e^{-\beta E_{m} 
}} \mathord{\left/ {\vphantom {{e^{-\beta E_{m} }} \Xi }} \right. 
\kern-\nulldelimiterspace} \Xi $ would be nearly equal to unity and zero for 
the ground state and excited states, respectively. This speculation seems to 
be appropriate because there generally exists the energy gap between the 
ground state and excited states for superconductors. Of course it is 
rigorously correct at zero temperature. Further discussion will be presented 
in Sec. III-D.

Accordingly, the dominant contribution in Eq. (\ref{eq43}) at low temperature comes 
from the case for 
${\left| {\Phi_{m}} \right\rangle}\!\!=\!\!{\left| {\Phi_{m'}} \right\rangle}\!\!=\!\!\rm{ground \,\, state}$. 
That is to say, Eq. (\ref{eq43}) is approximated at low temperature in the following form:
\begin{equation}
\label{eq45}
\int {\left\langle {\hat{{D}}^{(1)}\left( {{\rm {\bf r}}_{1} \zeta_{1} 
,{\rm {\bf r}}_{2} \zeta_{2} } \right)} \right\rangle_{\!\!H_{M} } 
\left\langle {\hat{{D}}^{(1)}\left( {{\rm {\bf r}}_{2} \zeta_{2} ,{\rm {\bf 
r}}_{1} \zeta_{1} } \right)} \right\rangle_{\!\!H_{M} } d^{3}r_{1} d^{3}r_{2} 
d\zeta_{1} d\zeta_{2} } \approx \sum\limits_{\mu_{\Phi_{m} } } {\left( 
{{\frac{e^{-\beta E_{m} }}{\Xi }}n_{\mu_{\Phi_{m} } }^{(1)} } \right)^{2}},
\end{equation}
where we use the orthonormality of the natural orbitals, and where the 
summation of the right-hand side is over the natural spin orbitals only for 
the ground state. Similarly, Eq. (\ref{eq44}) is approximated at low temperature as
\begin{equation}
\label{eq46}
\int {\left\langle {\hat{{D}}^{(1)}\left( {{\rm {\bf r}}_{1} \zeta_{1} 
,{\rm {\bf r}}_{1} \zeta_{1} } \right)} \right\rangle_{\!H_{M} } } 
d^{3}r_{1} d\zeta_{1} \approx \sum\limits_{\mu_{\Phi_{m} } } 
{{\frac{e^{-\beta E_{m} }}{\Xi }}n_{\mu_{\Phi_{m} } }^{(1)} } ,
\end{equation}
where the note on the summation is the same as that of Eq. (\ref{eq45}).

Furthermore, since the eigenfunction ${\left| {\Phi_{m}} \right\rangle }$ 
of Eq. (\ref{eq40}) is the single Slater determinant in terms of the quasiparticle, 
and since the eigenvalue $n_{\mu_{\Phi_{m} } }^{(1)} $ corresponds to the 
occupation number of the natural spin orbital in the single Slater 
determinant ${\left| {\Phi_{m}} \right\rangle }$\cite{2}, the value of $n_{\mu 
_{\Phi_{m} } }^{(1)} $ is necessarily equal to unity or zero. Thus, we have 
\begin{equation}
\label{eq47}
{\frac{e^{-\beta E_{m} }}{\Xi }}\,n_{\mu_{\Phi_{m} } }^{(1)} \,\approx 
\,\,1\,\,\,\,\,\mbox{or}\,\,\,\,\,0.
\end{equation}
Subtracting Eq. (\ref{eq46}) from Eq. (\ref{eq45}) on both sides, we have
\begin{eqnarray}
\label{eq48}
\!\!\!&&\!\!\!\!\!\int \!\!{\left\langle {\hat{{D}}^{(1)}\left( {{\rm {\bf r}}_{1} \zeta_{1} 
,{\rm {\bf r}}_{2} \zeta_{2} } \right)} \right\rangle_{\!\!H_{M} } \!\!
\left\langle {\hat{{D}}^{(1)}\left( {{\rm {\bf r}}_{2} \zeta_{2} ,{\rm {\bf 
r}}_{1} \zeta_{1} } \right)} \right\rangle_{\!\!H_{M} } \!\!\!d^{3}r_{1} d^{3}r_{2} 
d\zeta_{1} d\zeta_{2} } -\!\int \!{\left\langle {\hat{{D}}^{(1)}\left( {{\rm 
{\bf r}}_{1} \zeta_{1} ,{\rm {\bf r}}_{1} \zeta_{1} } \right)} 
\right\rangle_{\!H_{M} } } \!\!d^{3}r_{1} d\zeta_{1} \nonumber \\ 
&\approx& \sum\limits_{\mu_{\Phi_{m} } } {\left( {{\frac{e^{-\beta E_{m} 
}}{\Xi }}n_{\mu_{\Phi_{m} } }^{(1)} } \right)\left( {{\frac{e^{-\beta 
E_{m} }}{\Xi }}n_{\mu_{\Phi_{m} } }^{(1)} -1} \right)} . 
\end{eqnarray}
Due to Eq. (\ref{eq47}), the right-hand side of Eq. (\ref{eq48}) is approximately equal to 
zero at low temperature. We finally obtain
\begin{equation}
\label{eq49}
\!\! \int \!\!{\left\langle {\hat{{D}}^{(1)}\left( {{\rm {\bf r}}_{1} \zeta_{1} 
,{\rm {\bf r}}_{2} \zeta_{2} } \right)} \right\rangle_{\!\!H_{M} } \!\!
\left\langle {\hat{{D}}^{(1)}\left( {{\rm {\bf r}}_{2} \zeta_{2} ,{\rm {\bf 
r}}_{1} \zeta_{1} } \right)} \right\rangle_{\!\!H_{M} } \!\!d^{3}r_{1} d^{3}r_{2} 
d\zeta_{1} d\zeta_{2} } \approx \!\int\! {\left\langle {\hat{{D}}^{(1)}\left( 
{{\rm {\bf r}}_{1} \zeta_{1} ,{\rm {\bf r}}_{1} \zeta_{1} } \right)} 
\right\rangle_{\!\!H_{M} } } d^{3}r_{1} d\zeta_{1} .
\end{equation}
Thus, it is shown that the idempotent of the RDM1 approximately holds at low 
temperature even in the superconducting state. It should be noted that Eq. 
(\ref{eq49}) rigorously holds at zero temperature.
%
\subsection{Fluctuation of the particle number for superconductors at low temperature}
\label{secIII-C}
In this section, it is shown that the fluctuation of the particle number is 
equal to $O(N)$ when the OPSS takes nonzero value.

Under the condition that ${\rm {\bf r}}_{1} ,\,\zeta_{1} ={\rm {\bf r}}_{4} 
,\,\zeta_{4} $ and ${\rm {\bf r}}_{2} ,\,\zeta_{2} ={\rm {\bf r}}_{3} 
,\,\zeta_{3} $, the Bloch-de Dominicis theorem for the RDM2, i.e., Eq. 
(\ref{eq35}), is written as
\begin{eqnarray}
\label{eq50}
\left\langle {\psi^{\dag }({\rm {\bf r}}_{1} \zeta_{1} )\psi^{\dag 
}({\rm {\bf r}}_{2} \zeta_{2} )\psi ({\rm {\bf r}}_{2} \zeta_{2} )\psi 
({\rm {\bf r}}_{1} \zeta_{1} )} \right\rangle_{\!H_{M} } 
&\!=\!&\left\langle {\psi ^{\dag }({\rm {\bf r}}_{1} \zeta_{1} )\psi ({\rm {\bf r}}_{1} \zeta_{1} )} 
\right\rangle_{\!H_{M} } 
\left\langle {\psi^{\dag }({\rm {\bf r}}_{2} \zeta _{2} )\psi ({\rm {\bf r}}_{2} \zeta_{2} )} 
\right\rangle_{\!H_{M} } \nonumber \\ 
&\!-\!&\left\langle {\psi^{\dag }({\rm {\bf r}}_{1} \zeta_{1} )\psi ({\rm {\bf r}}_{2} \zeta_{2} )} 
\right\rangle_{\!H_{M} } 
\left\langle {\psi^{\dag }({\rm {\bf r}}_{2} \zeta_{2} )\psi ({\rm {\bf r}}_{1} \zeta_{1} )} 
\right\rangle_{\!H_{M} } \nonumber \\ 
&\!+\!&\left\langle {\psi^{\dag }({\rm {\bf r}}_{1} \zeta_{1} )\psi^{\dag }({\rm {\bf r}}_{2} \zeta_{2} )} 
\right\rangle_{\!H_{M} } 
\left\langle {\psi ({\rm {\bf r}}_{2} \zeta_{2} )\psi ({\rm {\bf r}}_{1} \zeta_{1} )} 
\right\rangle_{\!H_{M} }.
\end{eqnarray}
Using the anti-commutation relation of the field operators of electrons in 
the left-hand side, Eq. (\ref{eq50}) is rewritten as
\begin{eqnarray}
\label{eq51}
 \left\langle {\psi^{\dag }({\rm {\bf r}}_{1} \zeta_{1} )\psi ({\rm {\bf r}}_{1} \zeta_{1} )
\psi^{\dag }({\rm {\bf r}}_{2} \zeta_{2} )\psi ({\rm {\bf r}}_{2} \zeta_{2} )} \right\rangle_{\!H_{M} } 
&\!=\!&\left\langle {\psi ^{\dag }({\rm {\bf r}}_{1} \zeta_{1} )\psi ({\rm {\bf r}}_{2} \zeta_{2} )} 
\right\rangle_{\!H_{M} } \delta \left( {{\rm {\bf r}}_{1} -{\rm {\bf r}}_{2} 
} \right)\delta_{\zeta_{1} \zeta_{2} } \nonumber \\ 
&\!+\!&\left\langle {\psi^{\dag }({\rm {\bf r}}_{1} \zeta_{1} )\psi ({\rm {\bf r}}_{1} \zeta_{1} )} 
\right\rangle_{\!H_{M} } 
\left\langle {\psi^{\dag }({\rm {\bf r}}_{2} \zeta_{2} )\psi ({\rm {\bf r}}_{2} \zeta_{2} )} 
\right\rangle_{\!H_{M} } \nonumber \\ 
&\!-\!&\left\langle {\psi^{\dag }({\rm {\bf r}}_{1} \zeta_{1} )\psi ({\rm {\bf r}}_{2} \zeta_{2} )} 
\right\rangle_{\!H_{M} } 
\left\langle {\psi^{\dag }({\rm {\bf r}}_{2} \zeta_{2} )\psi ({\rm {\bf r}}_{1} \zeta_{1} )} 
\right\rangle_{\!H_{M} } \nonumber \\ 
&\!+\!&\left\langle {\psi^{\dag }({\rm {\bf r}}_{1} \zeta_{1} )\psi^{\dag }({\rm {\bf r}}_{2} \zeta_{2} )} 
\right\rangle_{\!H_{M} } \left\langle {\psi 
({\rm {\bf r}}_{2} \zeta_{2} )\psi ({\rm {\bf r}}_{1} \zeta_{1} )} 
\right\rangle_{\!H_{M} }.
\end{eqnarray}
Integrating both sides with respect to ${\rm {\bf r}}_{1} ,\,\,{\rm {\bf 
r}}_{2} $ and $\zeta_{1} ,\,\,\zeta_{2} $, we obtain 
\begin{eqnarray}
\label{eq52}
\left\langle {\hat{{N}}^{2}} \right\rangle_{H_{M} } 
&=&\left\langle {\hat{{N}}} \right\rangle_{H_{M} } 
+\left( {\left\langle {\hat{{N}}} \right\rangle_{H_{M} } } \right)^{2} \nonumber \\
&-&\int {\left\langle {\hat{{D}}^{(1)}\left( {{\rm {\bf r}}_{1} \zeta_{1} ,{\rm {\bf r}}_{2} 
\zeta_{2} } \right)} \right\rangle_{H_{M} } \left\langle 
{\hat{{D}}^{(1)}\left( {{\rm {\bf r}}_{2} \zeta_{2} ,{\rm {\bf r}}_{1} 
\zeta_{1} } \right)} \right\rangle_{H_{M} } d^{3}r_{1} d\zeta_{1} 
d^{3}r_{2} d\zeta_{2} } \nonumber \\ 
&+&\int {\left\langle {\psi^{\dag }({\rm {\bf r}}_{1} \zeta_{1} )\psi ^{\dag }({\rm {\bf r}}_{2} \zeta_{2} )} \right\rangle_{H_{M} } 
\left\langle {\psi ({\rm {\bf r}}_{2} \zeta_{2} )\psi ({\rm {\bf r}}_{1} 
\zeta_{1} )} \right\rangle_{H_{M} } } d^{3}r_{1} d\zeta_{1} d^{3}r_{2} 
d\zeta_{2} , 
\end{eqnarray}
where Eqs. (\ref{eq36}) and (\ref{eq39}) are used, and where $\hat{{N}}$ is given by
\begin{equation}
\label{eq53}
\hat{{N}}=\int {\psi^{\dag }({\rm {\bf r}}_{1} \zeta_{1} )\psi ({\rm {\bf 
r}}_{1} \zeta_{1} )d^{3}r_{1} d\zeta_{1} } .
\end{equation}
Using Eqs. (\ref{eq49}) and (\ref{eq53}), the third term on the right-hand side of Eq. (\ref{eq52}) 
can be approximated as
\begin{equation}
\label{eq54}
\int {\left\langle {\hat{{D}}^{(1)}\left( {{\rm {\bf r}}_{1} \zeta_{1} 
,{\rm {\bf r}}_{2} \zeta_{2} } \right)} \right\rangle_{H_{M} } 
\left\langle {\hat{{D}}^{(1)}\left( {{\rm {\bf r}}_{2} \zeta_{2} ,{\rm {\bf 
r}}_{1} \zeta_{1} } \right)} \right\rangle_{H_{M} } d^{3}r_{1} d\zeta_{1} 
d^{3}r_{2} d\zeta_{2} } \approx \left\langle {\hat{{N}}} \right\rangle 
_{H_{M} } ,
\end{equation}
at low temperature. Substituting Eq. (\ref{eq54}) into Eq. (\ref{eq52}), we finally get
\begin{equation}
\label{eq55}
\left\langle {\hat{{N}}^{2}} \right\rangle_{H_{M} } -\left( {\left\langle 
{\hat{{N}}} \right\rangle_{H_{M} } } \right)^{2}\approx \int {\left\langle 
{\psi^{\dag }({\rm {\bf r}}_{1} \zeta_{1} )\psi^{\dag }({\rm {\bf r}}_{2} 
\zeta_{2} )} \right\rangle_{H_{M} } \left\langle {\psi ({\rm {\bf r}}_{2} 
\zeta_{2} )\psi ({\rm {\bf r}}_{1} \zeta_{1} )} \right\rangle_{H_{M} } } 
d^{3}r_{1} d\zeta_{1} d^{3}r_{2} d\zeta_{2}.
\end{equation}
The physical quantity $\left\langle {\psi^{\dag }({\rm {\bf r}}_{1} \zeta 
_{1} )\psi^{\dag }({\rm {\bf r}}_{2} \zeta_{2} )} \right\rangle_{H_{M} } 
$ or $\left\langle {\psi ({\rm {\bf r}}_{2} \zeta_{2} )\psi ({\rm {\bf 
r}}_{1} \zeta_{1} )} \right\rangle_{H_{M} } $ is exactly the OPSS which 
has been mentioned in Sec. II-D. When the system is in the superconducting 
state, $\left\langle {\psi ({\rm {\bf r}}_{2} \zeta_{2} )\psi ({\rm {\bf 
r}}_{1} \zeta_{1} )} \right\rangle_{H_{M} } $ is given by Eq. (\ref{eq29}), and 
Eq. (\ref{eq55}) becomes
\begin{equation}
\label{eq56}
\left\langle {\hat{{N}}^{2}} \right\rangle_{H_{M} } -\left( {\left\langle 
{\hat{{N}}} \right\rangle_{H_{M} } } \right)^{2}\approx 2n_{\nu_{H_{M} 
}^{\max } }^{(2)} \int {\nu_{H_{M} }^{\max } \left( {{\rm {\bf r}}_{1} 
\zeta_{1} ,{\rm {\bf r}}_{2} \zeta_{2} } \right)^{\ast }\nu_{H_{M} 
}^{\max } \left( {{\rm {\bf r}}_{1} \zeta_{1} ,{\rm {\bf r}}_{2} \zeta_{2} 
} \right)} d^{3}r_{1} d\zeta_{1} d^{3}r_{2} d\zeta_{2} .
\end{equation}
Using the normalization of $\left| \nu_{H_{M}}^{\max} \right\rangle $ and applying Eqs. 
(\ref{eq10}) and (\ref{eq26}), we finally get 
\begin{equation}
\label{eq57}
\left\langle {\hat{{N}}^{2}} \right\rangle_{H_{M} } -\left( {\left\langle 
{\hat{{N}}} \right\rangle_{H_{M} } } \right)^{2}\approx \,O(N) .
\end{equation}
Thus, it is shown that the fluctuation of the particle number becomes $O(N)$ 
when the system is at low temperature in the superconducting state.
%
\subsection{Two kinds of fluctuations of the particle number in the superconducting state}
\label{secIII-D}
In the superconducting state, two kinds of fluctuations of the particle 
number can be generally observed. They are
\renewcommand{\labelenumi}{(\alph{enumi})} 
\begin{enumerate}
\item Statistical fluctuation of the particle number,
\item Quantum fluctuation of the particle number.
\end{enumerate}
The fluctuation (a) necessarily appears when the system is treated on the 
basis of the grand canonical ensemble \cite{74}. This is because the fluctuation 
(a) originates from the fact that the system is in the statistically-mixed 
state at the finite temperature. As is well known \cite{74}, the magnitude of the 
fluctuation (a) is $O(N)$. In the limit of zero temperature, the system is 
close to the pure state, which results in the disappearance of the 
statistical fluctuation of the particle number. The fluctuation (a) vanishes 
in the limit of zero temperature. Above the critical temperature of the 
superconductivity, the fluctuation (a) does not disappear and would become 
larger because the probabilities of the occurrence of states possessing the 
different numbers of particles increase with temperature. 

On the other hand, the fluctuation (b) appears only if the system is in the 
superconducting state, which has been shown in Sec. III-C. The fluctuation 
(b) originates from the spontaneous breaking of the gauge symmetry in the 
superconducting state, and does not from the statistically-mixed states. 
Therefore, the fluctuation (b) remains nonzero even at zero temperature, but 
vanishes at the critical temperature of the superconductivity. As the 
typical examples of this type of fluctuation, we can come up with the cases 
of the BCS ground state \cite{65,75} and the coherent state of the boson system 
\cite{76}. In reference to the discussion on the exciton \cite{77,78}, the latter case 
corresponds to the limiting case where the distance of particles forming the 
geminal is infinitely small and where the density of geminals is dilute. In 
both cases, the fluctuation of the particle number is shown to be $O(N)$ at 
zero temperature \cite{75,76,79}.

Here we shall give a comment on the discussion in Sec. III-C. Both 
fluctuations (a) and (b) should appear below the critical temperature. 
However, only the fluctuation (b) has been confirmed in Sec. III-C. 
This is not surprising because the fluctuation (a) becomes small at low 
temperature, and disappears in the limit of zero temperature. At low 
temperature, the fluctuation (a) can reasonably be disregarded compared to 
the fluctuation (b). In Sec. III-C, we have purposely focused on the case of 
the low temperature in order to evaluate the effects of the quantum 
fluctuation of the particle number. 

The fluctuations (a) and (b) are expected to be observed together in the 
superconducting state. The fluctuation (a) is an increasing function of the 
temperature regardless of above or below the critical temperature. On the 
other hand, the fluctuation (b) is a decreasing function of the temperature, 
and becomes zero at the critical temperature. The temperature dependence of 
the fluctuation obtained from the combined (a) and (b) would have a gentle 
slope below the critical temperature, while it would have a steep slope 
above the critical temperature. Using this fact, the critical temperature 
can be estimated quantitatively as the major changing point of the slope in 
the above-mentioned curve. That is to say, it is possible to estimate the 
critical temperature of the superconductivity by means of the fluctuation of 
the particle number. This is a strong merit of the PD functional theory 
which will be presented in the subsequent section.
%
\section{Pair-density functional theory for superconductors}
\label{secIV}
In the previous sections, it is shown that the fluctuation of the particle 
number may become an indication to judge whether the superconducting state 
appears or not. In Sec. IV-A, we shall show that the fluctuation of the 
particle number can be calculated directly by means of the PD and electron 
density. Considering this fact, the first thing we should do is to develop 
the first-principles theory where the PD and electron density are 
quantitatively reproduced. In Sec. IV-B, we present the theoretical 
framework of the PD functional theory for superconductors.
%
\subsection{Relation between the fluctuation of the particle number and PD}
\label{secIV-A}
In this subsection, it is shown that the fluctuation of the particle number 
can be calculated directly via the PD and electron density. The PD is 
defined as the diagonal elements of the RDM2. Using Eq. (\ref{eq1}), the operator of 
the PD, if it is denoted as $\hat{{\gamma }}^{(2)}\left( {{\rm {\bf r}}_{1} 
\zeta_{1} ,{\rm {\bf r}}_{2} \zeta_{2} ;{\rm {\bf r}}_{1} \zeta_{1} ,{\rm 
{\bf r}}_{2} \zeta_{2} } \right)$, is given by \cite{32} 
\begin{equation}
\label{eq58}
\hat{{\gamma }}^{(2)}\left( {{\rm {\bf r}}_{1} \zeta_{1} ,{\rm {\bf r}}_{2} 
\zeta_{2} ;{\rm {\bf r}}_{1} \zeta_{1} ,{\rm {\bf r}}_{2} \zeta_{2} } 
\right)={\frac{1}{2}}\psi^{\dag }({\rm {\bf r}}_{1} \zeta_{1} )\psi^{\dag 
}({\rm {\bf r}}_{2} \zeta_{2} )\psi ({\rm {\bf r}}_{2} \zeta_{2} )\psi 
({\rm {\bf r}}_{1} \zeta_{1} ).
\end{equation}
Using Eqs. (\ref{eq53}) and (\ref{eq58}), the operator $\hat{{N}}^{2}$ is expressed using 
$\hat{{\gamma }}^{(2)}\left( {{\rm {\bf r}}_{1} \zeta_{1} ,{\rm {\bf 
r}}_{2} \zeta_{2} ;{\rm {\bf r}}_{1} \zeta_{1} ,{\rm {\bf r}}_{2} \zeta 
_{2} } \right)$ and $\hat{{D}}^{(1)}\left( {{\rm {\bf r}}_{1} \zeta_{1} 
,{\rm {\bf r}}_{1} \zeta_{1} } \right)$. We have
\begin{equation}
\label{eq59}
\hat{{N}}^{2}=2\int\!\!\!\int {\hat{{\gamma }}^{(2)}\left( {{\rm {\bf 
r}}_{1} \zeta_{1} ,{\rm {\bf r}}_{2} \zeta_{2} ;{\rm {\bf r}}_{1} \zeta 
_{1} ,{\rm {\bf r}}_{2} \zeta_{2} } \right)d^{3}r_{1} d^{3}r_{2} d\zeta 
_{1} d\zeta_{2} } +\int {\hat{{D}}^{(1)}({\rm {\bf r}}\zeta ,{\rm {\bf 
r}}\zeta )d^{3}rd\zeta } ,
\end{equation}
where note that $\hat{{D}}^{(1)}\left( {{\rm {\bf r}}_{1} \zeta_{1} ,{\rm 
{\bf r}}_{1} \zeta_{1} } \right)$ is identical with the operator of the 
electron density. Therefore, the fluctuation of the particle number for the 
system with the statistical density matrix $\hat{{\rho }}_{H} $ is 
\begin{eqnarray}
\label{eq60}
\left\langle {\hat{{N}}^{2}} \right\rangle_{H} -\left\langle {\hat{{N}}} \right\rangle_{H}^{2} 
&=&2\int\!\!\!\int {\left\langle {\hat{{\gamma }}^{(2)}\left( {{\rm {\bf r}}_{1} \zeta_{1} ,{\rm {\bf r}}_{2} 
\zeta_{2} ;{\rm {\bf r}}_{1} \zeta_{1} ,{\rm {\bf r}}_{2} \zeta_{2} } \right)} 
\right\rangle_{H} d^{3}r_{1} d^{3}r_{2} d\zeta_{1} d\zeta_{2} } \nonumber \\
&+&\int {\left\langle {\hat{{D}}^{(1)}({\rm {\bf r}}\zeta ,{\rm {\bf r}}\zeta )} 
\right\rangle_{H} d^{3}rd\zeta }  
-\left\{ {\int {\left\langle {\hat{{D}}^{(1)}({\rm {\bf r}}\zeta ,{\rm {\bf r}}\zeta )} 
\right\rangle_{H} d^{3}rd\zeta } } \right\}^{2} . 
\end{eqnarray}
The fluctuation of the particle number can be calculated rigorously by means 
of the PD and electron density through Eq. (\ref{eq60}). In order to predict the 
fluctuation of the particle number, that is to say, in order to observe 
whether the OPSS appears in the system or not, we may just develop the 
first-principles theory in which both the PD and electron density are 
reproduced.

It should be noted that the electron density can be derived from the PD when 
the system is in the normal state \cite{61}, while it cannot be done in the 
superconducting state. There is no connection between the PD and electron 
density in the superconducting state. This is because the superconducting 
state does not satisfy the particle number conservation \cite{80}. In the 
superconducting state, we have to choose both the PD and electron density as 
the basic variables for describing the fluctuation of the particle number. 

%
\subsection{Pair-density functional theory}
\label{secIV-B}
In order to reproduce the fluctuation of the particle number, the PD 
functional theory seems to be most suitable because that for the normal 
state has been developed so far by many workers 
\cite{30,31,32,33,34,35,36,37,38,39,40,41,42,43,44,45,46,47,48,49,50,51,52,53,54,
55,56,57,58,59,60}, and because the 
findings and knowledge obtained by them may be useful for developing the 
theory for superconductors.

In a similar way to the zero-temperature PD functional theory \cite{32,33,34,
35,36,37,38,39,40}, the finite temperature PD functional theory can be formulated 
on the basis of so-called the `` Hohenberg-Kohn theorem''. It is composed of two theorems; 
one is the variational principle with respect to the PD and electron 
density, and another is the theorem concerning one-to-one correspondence 
between the correct density matrix and equilibrium densities. We shall give 
the proofs of them below.

Let us start with the Hamiltonian for superconductors. It includes the 
kinetic energy $\hat{{T}}$, electron-electron repulsive potential energy 
$\hat{{W}}_{1} $, external potential energy $\hat{{V}}$, and 
electron-electron attractive potential energy $\hat{{W}}_{2} $ which is 
mediated by the quasiparticle such as a phonon. We have 
\begin{equation}
\label{eq61}
\!\!\!\!\!\!\!\!\!\!\!\!\!\!\!\!\!\!\!\!\!\!\!\!\!\!\!\!\!\!\!\!\!\!\!\!\!\!\!\!\!\!\!\!\!\!\!\!
\!\!\!\!\!\!\!\!\!\!\!\!\!\!\!\!\!\!\!\!\!\!\!\!\!\!\!\!\!\!\!\!\!\!\!\!
\hat{{H}}=\hat{{T}}+\hat{{W}}_{1} +\hat{{W}}_{2} +\hat{{V}},
\end{equation}
with
\begin{equation}
\label{eq62}
\!\!\!\!\!\!\!\!\!\!\!\!\!\!\!\!\!\!\!\!\!\!\!\!\!\!\!\!\!\!\!\!\!\!\!\!\!\!\!\!\!\!\!\!\!\!\!\!
\!\!\!\!\!\!\!\!\!\!\!\!\!\!\!\!\!\!\!\!\!\!\!
\hat{{T}}=\sum\limits_\sigma {\int {\psi_{\sigma }^{\dag } ({\rm {\bf 
r}}){\frac{{\rm {\bf p}}^{2}}{2m}}\psi_{\sigma } ({\rm {\bf r}})d^{3}r} },
\end{equation}
\begin{equation}
\label{eq63}
\hat{{W}}_{1} ={\frac{1}{2}}\sum\limits_\sigma {\sum\limits_{{\sigma }'} 
{\int\!\!\!\int {\psi_{\sigma }^{\dag } ({\rm {\bf r}})\psi_{{\sigma 
}'}^{\dag } ({\rm {\bf {r}'}}){\frac{e^{2}}{\left| {{\rm {\bf r}}-{\rm {\bf 
{r}'}}} \right|}}\psi_{{\sigma }'} ({\rm {\bf {r}'}})\psi_{\sigma } ({\rm 
{\bf r}})d^{3}rd^{3}{r}'} } } ,
\end{equation}
\begin{equation}
\label{eq64}
\hat{{W}}_{2} ={\frac{1}{2}}\sum\limits_\sigma {\sum\limits_{{\sigma }'} 
{\int\!\!\!\int {\psi_{\sigma }^{\dag } ({\rm {\bf r}})\psi_{{\sigma 
}'}^{\dag } ({\rm {\bf {r}'}})w({\rm {\bf r}},\,{\rm {\bf {r}'}})\psi 
_{{\sigma }'} ({\rm {\bf {r}'}})\psi_{\sigma } ({\rm {\bf 
r}})d^{3}rd^{3}{r}'} } } ,
\end{equation}
\begin{equation}
\label{eq65}
\!\!\!\!\!\!\!\!\!\!\!\!\!\!\!\!\!\!\!\!\!\!\!\!\!\!\!\!\!\!\!\!\!\!\!\!\!\!\!\!\!\!\!\!\!\!\!\!
\!\!\!\!\!\!\!\!\!\!\!\!\!\!\!\!\!\!\!\!\!\!\!
\hat{{V}}=\sum\limits_\sigma {\int {v({\rm {\bf r}})\psi_{\sigma }^{\dag } 
({\rm {\bf r}})\psi_{\sigma } ({\rm {\bf r}})d^{3}r} } ,
\end{equation}
where $\psi_{\sigma } ({\rm {\bf r}})$ is the field operator of the 
electron with spin $\sigma $, and where $w({\rm {\bf r}},\,{\rm {\bf 
{r}'}})$ and $v({\rm {\bf r}})$ are the attractive potential and external 
potential, respectively. 

We construct the PD functional theory in accordance with the way of the 
extended constrained-search theory \cite{81,82,83,84,85}\cite{25}. The PD and 
electron density are chosen as the basic variables which determine the properties 
of the equilibrium state of the system. The statistical averages of them with 
respect to the statistical density matrix $\hat{{\rho }}$ are, respectively, 
written as
\begin{eqnarray}
\label{eq66}
&&\gamma^{(2)}\left( {{\rm {\bf r}}\zeta ,{\rm {\bf {r}'}}{\zeta }';{\rm 
{\bf r}}\zeta ,{\rm {\bf {r}'}}{\zeta }'} \right) \nonumber \\
&\!\!\!\!\!=&
\mbox{Tr}\left\{ {\hat{{\rho }}\hat{{\gamma }}^{(2)}\left( {{\rm {\bf r}}\zeta ,{\rm {\bf {r}'}}
{\zeta }';{\rm {\bf r}}\zeta ,{\rm {\bf {r}'}}{\zeta }'} \right)} \right\} \nonumber \\ 
&\!\!\!\!\!=&{\frac{1}{2}}\sum\limits_{\sigma_{1} } {\sum\limits_{\sigma_{2} } 
{\sum\limits_{\sigma_{3} } {\sum\limits_{\sigma_{4} } {\mbox{Tr}\left\{ 
{\hat{{\rho }}\psi_{\sigma_{1} }^{\dag } ({\rm {\bf r}})\psi_{\sigma_{2} 
}^{\dag } ({\rm {\bf {r}'}})\psi_{\sigma_{3} } ({\rm {\bf {r}'}})\psi 
_{\sigma_{4} } ({\rm {\bf r}})} \right\}} } } } \chi_{\sigma_{1} } (\zeta 
)\chi_{\sigma_{4} } (\zeta )\chi_{\sigma_{2} } ({\zeta }')\chi_{\sigma 
_{3} } ({\zeta }'), 
\end{eqnarray}
and
\begin{eqnarray}
\label{eq67}
n\left( {{\rm {\bf r}}\zeta } \right)
&=&\mbox{Tr}\left\{ {\hat{{\rho }}\hat{{n}}\left( {{\rm {\bf r}}\zeta } \right)} \right\} 
\nonumber \\ 
&=&\sum\limits_{\sigma_{1} } {\sum\limits_{\sigma_{2} } {\mbox{Tr}\left\{ 
{\hat{{\rho }}\psi_{\sigma_{1} }^{\dag } ({\rm {\bf r}})\psi_{\sigma_{2} 
} ({\rm {\bf r}})} \right\}\chi_{\sigma_{1} } (\zeta )\chi_{\sigma_{2} } 
(\zeta )} } . 
\end{eqnarray}
Here we define the universal energy functional that are independent of the 
external field:
\begin{equation}
\label{eq68}
F\left[ {\gamma^{(2)},n} \right]=\mathop{\mbox{Min}}\limits_{\hat{\rho}\to 
\left( {\gamma^{(2)},n} \right)} \mbox{Tr}\left\{ {\hat{{\rho }}\left( 
{\hat{{T}}+\hat{{W}}_{1} +\hat{{W}}_{2} } \right)+{\frac{1}{\beta 
}}\hat{{\rho }}\ln \hat{{\rho }}} \right\},
\end{equation}
where the right-hand side means that the minimum value of the statistical 
average of $\hat{{T}}+\hat{{W}}_{1} +\hat{{W}}_{2} $ plus entropy-related 
term is searched by varying statistical density matrices within the set of 
those that yield prescribed $\gamma^{(2)}$ and $n$. The last term in the 
curly bracket on the right-hand side is the entropy term, the operator of 
which is given by $\hat{{S}}=-k_{B} \hat{{\rho }}\ln \hat{{\rho }}$.

We shall discuss the Hohenberg-Kohn theorem for the present PD functional 
theory. With the use of Eq. (\ref{eq68}), the variational principle with respect to 
the PD and electron density can be derived by rewriting Gibbs's variational 
principle for the equilibrium density matrix \cite{86}. First the grand-potential 
functional in terms of the density matrix is defined as follows:
\begin{equation}
\label{eq69}
\Omega \left[ {\hat{{\rho }}} \right]=\mbox{Tr}\left\{ {\hat{{\rho }}\left( 
{\hat{{H}}-\mu \hat{{N}}} \right)+{\frac{1}{\beta }}\hat{{\rho }}\ln 
\hat{{\rho }}} \right\},
\end{equation}
where $\hat{{N}}$ is the operator of the particle number which is given by 
$\hat{{N}}=\int {\hat{{n}}({\rm {\bf r}},\zeta )d^{3}rd\zeta } $. Gibbs's 
variational principle says that the above functional $\Omega \left[ 
{\hat{{\rho }}} \right]$ takes the minimum value at the correct density 
matrix $\hat{{\rho }}_{0} $ and the corresponding value $\Omega \left[ 
{\hat{{\rho }}_{0} } \right]$ is equal to the correct grand potential, i.e., 
the grand potential of the equilibrium state $\Omega_{0} $. Namely we have
\begin{equation}
\label{eq70}
\Omega_{0} =\mathop{\mbox{Min}}\limits_{\hat{{\rho }}} \Omega \left[ {\hat{{\rho 
}}} \right]=\Omega \left[ {\hat{{\rho }}_{0} } \right],
\end{equation}
where $\hat{{\rho }}_{0} $ is given by \cite{86}
\begin{equation}
\label{eq71}
\hat{{\rho }}_{0} ={\frac{e^{-\beta \left( {\hat{{H}}-\mu \hat{{N}}} 
\right)}}{\Xi }},
\end{equation}
with $\Xi =\mbox{Tr}\left\{ {e^{-\beta \left( {\hat{{H}}-\mu \hat{{N}}} \right)}} \right\}$.  
Equation (\ref{eq70}) is formally divided into two-step variations such that
\begin{eqnarray}
\label{eq72}
\!\!\!\!\!\!\!\!\Omega_{0} 
&=&\mathop{\mbox{Min}}\limits_{\gamma^{(2)},\,n} \left\{ 
{\mathop{\mbox{Min}}\limits_{\hat{{\rho }}\to \left( {\gamma^{(2)},\,n} \right)} 
\Omega \left[ {\hat{{\rho }}} \right]} \right\} \nonumber \\ 
&=&\mathop{\mbox{Min}}\limits_{\gamma^{(2)},\,n} \left[ 
{\mathop{\mbox{Min}}\limits_{\hat{{\rho }}\to \left( {\gamma^{(2)},\,n} \right)} 
\left[ {\mbox{Tr}\left\{ {\hat{{\rho }}\left( {\hat{{T}}+\hat{{W}}_{1} 
+\hat{{W}}_{2} } \right)+{\frac{1}{\beta }}\hat{{\rho }}\ln \hat{{\rho }}} 
\right\}+\mbox{Tr}\left\{ {\hat{{\rho }}\left( {\hat{{V}}-\mu \hat{{N}}} \right)} 
\right\}} \right]} \right] \nonumber \\ 
&=&\mathop{\mbox{Min}}\limits_{\gamma^{(2)},\,n} \left[ {F\left[ {\gamma 
^{(2)},\,n} \right]+\int {\left\{ {v({\rm {\bf r}})-\mu } \right\}n({\rm {\bf 
r}},\zeta )d^{3}rd\zeta } } \right], 
\end{eqnarray}
where Eqs. (\ref{eq65}), (\ref{eq67}) and (\ref{eq68}) are used. If we define the grand potential 
functional which is given by
\begin{equation}
\label{eq73}
\Omega_{v-\mu } \left[ {\gamma^{(2)},\,n} \right]=F\left[ {\gamma 
^{(2)},\,n} \right]+\int {\left\{ {v({\rm {\bf r}})-\mu } \right\}n({\rm 
{\bf r}},\zeta )d^{3}rd\zeta } ,
\end{equation}
then Eq. (\ref{eq72}) is rewritten as
\begin{equation}
\label{eq74}
\Omega_{0} =\mathop{\mbox{Min}}\limits_{\gamma^{(2)},\,n} \Omega_{v-\mu } 
\left[ {\gamma^{(2)},\,n} \right].
\end{equation}
Compared the first line of the right-hand side of Eq. (\ref{eq72}) with the third 
line, the functional $\Omega_{v-\mu } \left[ {\gamma^{(2)},\,n} \right]$ 
is written as
\begin{equation}
\label{eq75}
\Omega_{v-\mu } \left[ {\gamma^{(2)},\,n} 
\right]=\mathop{\mbox{Min}}\limits_{\hat{{\rho }}\to \left( {\gamma^{(2)},\,n} 
\right)} \Omega \left[ {\hat{{\rho }}} \right].
\end{equation}
The value of $\Omega_{v-\mu } \left[ {\gamma^{(2)},\,n} \right]$ 
corresponds to the grand potential at the minimum point within the 
restricted set of density matrices that yield the prescribed $\gamma^{(2)}$ 
and $n$. Therefore, Eq. (\ref{eq74}) means that the global minimum point is searched 
within the set of local minimum points searched by Eq. (\ref{eq75}). Since such a 
global minimum point gives the correct grand potential $\Omega_{0} $, and 
using Eq. (\ref{eq70}), the PD and electron density that are found via Eq. (\ref{eq74}) 
corresponds to those that are calculated from the correct density matrix 
$\hat{{\rho }}_{0} $. They are exactly the PD and electron density for the 
equilibrium state of the system, which are hereafter denoted as $\gamma 
_{0}^{(2)} $ and $n_{0} $. Thus, Eq. (\ref{eq74}) represents the variational 
principle with respect to the PD and electron density. The results are 
summarized as follows:
\begin{equation}
\label{eq76}
\Omega_{0} =\mathop{\mbox{Min}}\limits_{\gamma^{(2)},\,n} \Omega_{v-\mu } 
\left[ {\gamma^{(2)},\,n} \right]=\Omega_{v-\mu } \left[ {\gamma 
_{0}^{(2)} ,\,n_{0} } \right],
\end{equation}
with
\begin{equation}
\label{eq77}
\begin{array}{l}
 \gamma_{0}^{(2)} \left( {{\rm {\bf r}}\zeta ,{\rm {\bf {r}'}}{\zeta 
}';{\rm {\bf r}}\zeta ,{\rm {\bf {r}'}}{\zeta }'} \right)=\mbox{Tr}\left\{ 
{\hat{{\rho }}_{0} \hat{{\gamma }}^{(2)}\left( {{\rm {\bf r}}\zeta ,{\rm 
{\bf {r}'}}{\zeta }';{\rm {\bf r}}\zeta ,{\rm {\bf {r}'}}{\zeta }'} \right)} 
\right\}, \\ 
 n_{0} \left( {{\rm {\bf r}}\zeta } \right)=\mbox{Tr}\left\{ {\hat{{\rho 
}}_{0} \hat{{n}}\left( {{\rm {\bf r}}\zeta } \right)} \right\}. \\ 
 \end{array}
\end{equation}

Next we discuss the theorem for one-to-one correspondence between the 
correct density matrix and equilibrium densities. The universal energy 
functional at the equilibrium densities is given by
\begin{eqnarray}
\label{eq78}
F\left[ {\gamma_{0}^{(2)} ,\,n_{0} } \right]
&=&\!\!\!
\mathop{\mbox{Min}}\limits_{\hat{{\rho }}\to \left( {\gamma_{0}^{(2)} ,\,n_{0} } \right)} 
\!\!\mbox{Tr}\left\{ {\hat{{\rho }}\left( 
{\hat{{T}}+\hat{{W}}_{1} +\hat{{W}}_{2} } \right)+{\frac{1}{\beta 
}}\hat{{\rho }}\ln \hat{{\rho }}} \right\} \nonumber \\ 
&=&
\mbox{Tr}\left\{ {\hat{{\rho }}_{\min } \left( {\hat{{T}}+\hat{{W}}_{1} 
+\hat{{W}}_{2} } \right)+{\frac{1}{\beta }}\hat{{\rho }}_{\min } \ln 
\hat{{\rho }}_{\min } } \right\}, 
\end{eqnarray}
where the searched density matrix of the right-hand side is referred to as 
$\hat{{\rho }}_{\min } $, and where note that $\hat{{\rho }}_{\min } $ 
yields $\gamma_{0}^{(2)} $ and $n_{0} $. Considering Eq. (\ref{eq70}), the 
following relation holds:
\begin{equation}
\label{eq79}
\begin{array}{l}
\mbox{Tr}\left\{ {\hat{{\rho }}_{0} \left( {\hat{{T}}+\hat{{W}}_{1} +\hat{{W}}_{2} 
+\hat{{V}}-\mu \hat{{N}}} \right)+\displaystyle{\frac{1}{\beta }}
\hat{{\rho }}_{0} \ln \hat{{\rho }}_{0} } \right\} \\ 
 \le \mbox{Tr}\left\{ {\hat{{\rho }}_{\min } \left( {\hat{{T}}+\hat{{W}}_{1} 
+\hat{{W}}_{2} +\hat{{V}}-\mu \hat{{N}}} \right)+
\displaystyle{\frac{1}{\beta }}\hat{{\rho }}_{\min } \ln \hat{{\rho }}_{\min } } \right\}. \\ 
 \end{array}
\end{equation}
Since both $\hat{{\rho }}_{0} $ and $\hat{{\rho }}_{\min } $ yield the 
correct densities $\gamma_{0}^{(2)} $ and $n_{0} $, it follows that
\begin{equation}
\label{eq80}
\mbox{Tr}\left\{ {\hat{{\rho }}_{0} \left( {\hat{{V}}-\mu \hat{{N}}} 
\right)} \right\}\mbox{=Tr}\left\{ {\hat{{\rho }}_{\min } \left( 
{\hat{{V}}-\mu \hat{{N}}} \right)} \right\}=\int {\left\{ {v({\rm {\bf 
r}})-\mu } \right\}n_{0} ({\rm {\bf r}},\zeta )d^{3}rd\zeta } .
\end{equation}
Substitution of Eq. (\ref{eq80}) into Eq. (\ref{eq79}) leads to
\begin{equation}
\label{eq81}
\begin{array}{l}
 \mbox{Tr}\left\{ {\hat{{\rho }}_{0} \left( {\hat{{T}}+\hat{{W}}_{1} 
+\hat{{W}}_{2} } \right)+\displaystyle{\frac{1}{\beta }}\hat{{\rho }}_{0} \ln \hat{{\rho 
}}_{0} } \right\} \\ 
 \le \mbox{Tr}\left\{ {\hat{{\rho }}_{\min } \left( {\hat{{T}}+\hat{{W}}_{1} 
+\hat{{W}}_{2} } \right)+\displaystyle{\frac{1}{\beta }}\hat{{\rho }}_{\min } \ln 
\hat{{\rho }}_{\min } } \right\}. \\ 
 \end{array}
\end{equation}
The right-hand side is coincident with $F\left[ {\gamma_{0}^{(2)} 
,\,\,n_{0} } \right]$ from Eq. (\ref{eq78}), which deduces that only an equal sign 
is satisfied in Eq. (\ref{eq81}). Namely we have
\begin{equation}
\label{eq82}
\begin{array}{l}
 \mbox{Tr}\left\{ {\hat{{\rho }}_{0} \left( {\hat{{T}}+\hat{{W}}_{1} 
+\hat{{W}}_{2} } \right)+\displaystyle{\frac{1}{\beta }}\hat{{\rho }}_{0} \ln \hat{{\rho 
}}_{0} } \right\} \\ 
 =\mbox{Tr}\left\{ {\hat{{\rho }}_{\min } \left( {\hat{{T}}+\hat{{W}}_{1} 
+\hat{{W}}_{2} } \right)+\displaystyle{\frac{1}{\beta }}\hat{{\rho }}_{\min } \ln 
\hat{{\rho }}_{\min } } \right\}. \\ 
 \end{array}
\end{equation}
Using Eqs. (\ref{eq80}) and (\ref{eq82}), only an equal sign holds also in Eq. (\ref{eq79}). The 
left-hand side of Eq. (\ref{eq79}) is exactly the correct grand potential $\Omega 
_{0} $. Considering Gibbs's variational theorem (\ref{eq70}), we finally get
\begin{equation}
\label{eq83}
\hat{{\rho }}_{\min } =\hat{{\rho }}_{0} .
\end{equation}
It follows that the correct density matrix $\hat{{\rho }}_{0} 
\,\,\,(=\hat{{\rho }}_{\min } )$ is uniquely determined by the correct 
densities $\gamma_{0}^{(2)} $ and $n_{0} $ via Eq. (\ref{eq78}), and vice versa due 
to Eq. (\ref{eq77}). Thus, it is proved that one-to-one correspondence between the 
density matrix $\hat{{\rho }}_{0} $ and densities $\gamma_{0}^{(2)} $, 
$n_{0} $.

The most crucial theorems of the PD functional theory have successfully been 
derived. These theorems enable us to develop the concrete scheme for 
calculating the PD and electron density of the equilibrium state. This would 
be the next issue to be tackled.
%

\subsection{Discussions}
\label{secIV-C}
As mentioned in Sec. III, the fluctuation of the particle number is a 
possible indication of the superconducting state, and accordingly the 
critical temperature of the superconductivity may be evaluated via the 
temperature dependence of such a fluctuation. This seems to be one of the 
strong merits of the PD functional theory presented in the previous section.

In addition to the estimation of the critical temperature, the present 
theory has another merit about the description of the superconducting state. 
As shown below, it can predict the density of particles which form the 
geminal ${\left| {\nu_{H_{M}}^{\max}} \right\rangle }$. In other words, the spatial 
distribution of the OPSS can be obtained from the present PD functional 
theory.

As shown in Sec. III-B, the idempotent of the RDM1 holds at low 
temperature. This idempotent can also be written in the other form:
\begin{equation}
\label{eq84}
\int {\left\langle {\hat{{D}}^{(1)}\left( {{\rm {\bf r}}_{1} \zeta_{1} 
,{\rm {\bf r}}_{2} \zeta_{2} } \right)} \right\rangle_{\!H_{M} } 
\left\langle {\hat{{D}}^{(1)}\left( {{\rm {\bf r}}_{2} \zeta_{2} ,{\rm {\bf 
r}}_{1} \zeta_{1} } \right)} \right\rangle_{\!H_{M} } d^{3}r_{2} d\zeta_{2} 
} \approx \left\langle {\hat{{D}}^{(1)}\left( {{\rm {\bf r}}_{1} \zeta_{1} 
,{\rm {\bf r}}_{1} \zeta_{1} } \right)} \right\rangle_{\!H_{M} } .
\end{equation}
Integrating both sides of Eq. (\ref{eq50}) with respect to ${\rm {\bf r}}_{2} $ and 
$\zeta_{2} $, we have
\begin{eqnarray}
\label{eq85}
&&
\int {\left\langle {\psi^{\dag }({\rm {\bf r}}_{1} \zeta_{1} )\psi^{\dag 
}({\rm {\bf r}}_{2} \zeta_{2} )\psi ({\rm {\bf r}}_{2} \zeta_{2} )\psi 
({\rm {\bf r}}_{1} \zeta_{1} )} \right\rangle_{H_{M} } } d^{3}r_{2} d\zeta 
_{2} \nonumber \\ 
&=&\left\langle {\psi^{\dag }({\rm {\bf r}}_{1} \zeta_{1} )\psi ({\rm {\bf r}}_{1} \zeta_{1} )} 
\right\rangle _{H_{M} } \int {\left\langle {\psi^{\dag }({\rm {\bf r}}_{2} \zeta_{2} )
\psi ({\rm {\bf r}}_{2} \zeta_{2} )} \right\rangle_{H_{M} } d^{3}r_{2} d\zeta_{2} } \nonumber \\ 
&-&\int {\left\langle {\psi ^{\dag }({\rm {\bf r}}_{1} \zeta_{1} )\psi ({\rm {\bf r}}_{2} \zeta_{2} )} 
\right\rangle_{H_{M} } \left\langle {\psi^{\dag }({\rm {\bf r}}_{2} \zeta _{2} )
\psi ({\rm {\bf r}}_{1} \zeta_{1} )} \right\rangle_{H_{M} } } d^{3}r_{2} d\zeta_{2} \nonumber \\ 
&+&\int {\left\langle {\psi ^{\dag }({\rm {\bf r}}_{1} \zeta_{1} )
\psi^{\dag }({\rm {\bf r}}_{2} \zeta _{2} )} \right\rangle_{H_{M} } \left\langle 
{\psi ({\rm {\bf r}}_{2} \zeta _{2} )\psi ({\rm {\bf r}}_{1} \zeta_{1} )} \right\rangle_{H_{M} } 
d^{3}r_{2} d\zeta_{2} }.
\end{eqnarray}
Substituting Eq. (\ref{eq84}) into Eq. (\ref{eq85}), and using Eqs. (\ref{eq23}), (\ref{eq29}) and (\ref{eq39}), Eq. 
(\ref{eq85}) is rewritten as
\begin{eqnarray}
\label{eq86}
&&2\!\int \!\!{\left\langle {\hat{{\gamma }}^{(2)}\left( {{\rm {\bf r}}_{1} \zeta 
_{1} {\rm {\bf r}}_{2} \zeta_{2} ;{\rm {\bf r}}_{1} \zeta_{1} {\rm {\bf 
r}}_{2} \zeta_{2} } \right)} \right\rangle_{H_{M} } } d^{3}r_{2} d\zeta 
_{2} -\left\{ {\left\langle {\hat{{N}}} \right\rangle_{H_{M} } -1} 
\right\}\left\langle {\hat{{n}}\left( {{\rm {\bf r}}_{1} \zeta_{1} } 
\right)} \right\rangle_{H_{M} } \nonumber \\ 
&=&2n_{\nu^{\max }}^{(2)} \int {\nu^{\max }({\rm {\bf r}}_{2} \zeta_{2} 
{\rm {\bf r}}_{1} \zeta_{1} )^{\ast }\nu^{\max }({\rm {\bf r}}_{1} \zeta 
_{1} {\rm {\bf r}}_{2} \zeta_{2} )d^{3}r_{2} d\zeta_{2} }. 
\end{eqnarray}
The integral of the right-hand side represents the density of particles 
forming geminals in the superconducting state. Thus, the spatial 
distribution of the OPSS can be obtained by means of the PD and electron 
density at low temperature. This is also a useful information for 
understanding the superconducting state as well as the critical temperature. 
%
\section{Concluding remarks}
\label{secV}
In this paper, we show that the fluctuation of the particle number is a 
possible indication of whether the superconducting state appears or not in 
the system. Specifically, the quantum fluctuation of the particle number 
becomes $O(N)$ when the OPSS appears in the system. Using this result, the 
critical temperature of the superconductivity can be evaluated in principle 
as a bending point in the temperature-dependence curve of the fluctuation of 
the particle number. 

As the practical scheme for calculating the fluctuation of the particle 
number, we also present the theoretical framework of the finite-temperature 
PD functional theory. This PD functional theory can evaluate not only the 
fluctuation of the particle number but also the density of particles forming 
geminals of the BEC. The latter corresponds to the spatial distribution of 
the OPSS, which also characterizes the superconducting state of the system 
as well as the critical temperature.

Let us discuss the meanings of the present PD functional theory from the 
viewpoint of the basic variables describing the superconducting state. In 
this paper, we present the theory where the fluctuation of the particle 
number, i.e., diagonal elements of the RDM2, are chosen as the basic 
variables, instead of choosing the off-diagonal elements of the RDM2 which 
have been used in the conventional theories for superconductors \cite{9,10,25}. 
In the superconducting state, the fluctuation of the particle number is 
closely related to the OPSS, which is shown in Eq. (\ref{eq55}). This means that the 
diagonal and off-diagonal elements of the RDM2 are related to each other in 
the superconducting state. Since the RDM2 is dependent on the 
superconducting state as shown in Eqs. (\ref{eq2}) and (\ref{eq23}), the diagonal and 
off-diagonal elements of the RDM2 are correlated mutually via the 
superconducting state. This is what we find in the present work. Although 
only the off-diagonal elements of the RDM2 have previously been adopted as 
the basic variables, the indication of the superconducting state necessarily 
appears also in the diagonal elements of the RDM2. This is exactly the 
quantum fluctuation of the particle number, which is the basic variable 
chosen in the present PD functional theory.

The next step is to construct the concrete scheme for calculating the PD and 
electron density on the basis of theorems of the present PD functional 
theory, and then to confirm the reproducibility of them through actual 
calculations. This will be done in near future.
\begin{acknowledgments}
This work was partially supported by Grant-in-Aid for Scientific Research 
(No. 26400354 and No. 26400397) of Japan Society for the Promotion of 
Science.
\end{acknowledgments}
\newpage


\begin{thebibliography}{00}
\bibitem{5}
The order parameter of the superconducting state is reviewed in Sec. II 
together with the discussion on the properties of two-particle states which 
forms the BEC.
\bibitem{6}
The representative first-principles theories for superconductors are the 
Bogoliubov-de Gennes theory \cite{7,8} and density functional theory \cite{9,10}. The 
explanation of the latter is followed in the text.
\bibitem{62}
This assumption is reasonable, because the superconductors usually have 
one kind of order parameters.
\bibitem{63}
It seems that electron pairings may lower the total energy of the 
system as can be seen in the BCS theory \cite{64,65}. If another kind of geminals 
belonging to type (b) exists, the occupation number of this geminal also 
becomes $O(N)$.
\bibitem{72}
The Bloch-de Dominicis theorem is sometimes called Wick's theorem at 
the finite temperature.
\bibitem{73}
The eigenfunction for the RDM1 is called the natural spin orbital in 
the field of the quantum chemistry \cite{61}.
\bibitem{79}
As is well known \cite{65,75}, the fluctuation of the particle number for 
the BCS ground state is proportional to the gap parameter at zero 
temperature and to the density of states at the Fermi level. Since the 
density of states is proportional to the volume of the system, the value of 
the above fluctuation is $O(N)$. Similarly, the fluctuation of the particle 
number for the coherent state of the boson system is rigorously equal to the 
number of the constituent boson \cite{76}.
\bibitem{80}
As shown in Sec. II, the OPSS is given by the average of $\psi ({\rm 
{\bf r}}\zeta )\psi ({\rm {\bf {r}'}}{\zeta }')$. It is obvious that the 
particle number conservation is broken when this order parameter has nonzero 
value.
\end{thebibliography}
\end{document}